%% file: main.tex
\documentclass[letterpaper,11pt]{article}

\usepackage{fullpage}
\usepackage{newclude}
\usepackage{amssymb}
\usepackage{amsmath}
\usepackage{amsthm}
\usepackage{algorithmicx}
\usepackage{algpseudocode}
\usepackage{hyperref}
\usepackage[margin=1in]{geometry}
\usepackage{mathptmx}
\usepackage[T1]{fontenc}
\usepackage{color}
\usepackage{comment}
\usepackage{fixltx2e}

\theoremstyle{plain}
\newtheorem{theorem}{Theorem}[section]
\newtheorem{lemma}[theorem]{Lemma}
\newtheorem{corollary}[theorem]{Corollary}

\newtheorem{claim}[theorem]{Claim}

\theoremstyle{definition}
\newtheorem{definition}[theorem]{Definition}

\theoremstyle{remark}

\DeclareMathAlphabet{\mathcal}{OMS}{cmsy}{m}{n}

\DeclareMathOperator{\st}{:}
\DeclareMathOperator{\CDF}{CDF}
\newcommand{\Rbb}{\mathbb{R}}
\newcommand{\Zbb}{\mathbb{Z}}
\newcommand{\eps}{\varepsilon}
\newcommand{\set}[1]{\left\{#1\right\}}
\newcommand{\Prb}[2]{\mathrm{Pr}_{#1}\left[#2\right]}
\newcommand{\Exp}[2]{\mathrm{E}_{#1}\left[#2\right]}
\newcommand{\EMD}{\mathrm{EMD}}

\newcommand{\Ac}{\mathcal{A}}

\newcommand{\Gc}{\mathcal{G}}
\newcommand{\Ic}{\mathcal{I}}
\newcommand{\Ec}{\mathcal{E}}

\newcommand{\Mc}{\mathcal{M}}

\DeclareMathOperator{\supp}{supp}
\DeclareMathOperator{\argmin}{argmin}

\newcommand{\poly}{\mathrm{poly}}

\newcommand{\Ball}{\mbox{Ball}}

\def\R{\mathbb{R}}

\def\eps{\epsilon}
\newcommand{\norm}[1]{\lVert#1\rVert}

\newcommand{\normm}[2]{\lVert#2\rVert_{#1}}

\newcommand{\wh}{\widehat}
\newcommand{\abs}[1]{\left|#1\right|}
\newcommand{\floor}[1]{\left\lfloor#1\right\rfloor}

\newcommand{\todo}[2]{}
\newcommand{\ir}[1]{\todo{blue}{IR: <<#1>>}}
\newcommand{\pji}[1]{\todo{red}{PI: <<#1>>}}

\MakeRobust{\Call}
\algtext*{EndWhile}
\algtext*{EndFor}
\algtext*{EndIf}
\algtext*{EndFunction}

\begin{document}

\begin{titlepage}

    \title{Nearly-optimal bounds for sparse recovery in generic norms, with applications to $k$-median sketching}
    \author{Arturs Backurs \and Piotr Indyk \and Eric Price \and Ilya Razenshteyn \and David P. Woodruff}
    \maketitle
\thispagestyle{empty}
\begin{abstract}
We initiate the study of trade-offs between sparsity and the number of measurements in sparse recovery schemes for {\em generic} norms. Specifically, for a norm $\|\cdot\|$, sparsity parameter $k$, approximation factor $K>0$, and probability of failure $P>0$, we ask: what is the minimal value of $m$ so that there is a distribution over $m \times n$ matrices $A$  with the property that for any~$x$, given $Ax$, we can recover a $k$-sparse approximation to $x$ in the given norm with probability at least $1-P$? We give a partial answer to this problem, by showing that for norms that admit efficient linear sketches, the optimal number of measurements $m$ is closely related to the {\em doubling dimension} of the metric induced by the norm $\|\cdot\|$ on the set of all $k$-sparse vectors. By applying our result to specific norms, we cast known measurement bounds in our general framework (for the $\ell_p$ norms, $p \in [1,2]$) as well as provide new, measurement-efficient schemes (for the Earth-Mover Distance norm). The latter result directly implies more succinct linear sketches for the well-studied planar {\em $k$-median clustering} problem. Finally, our lower bound for the doubling dimension of the EMD norm enables us to address the open question of [Frahling-Sohler, STOC'05] about the space complexity of clustering problems in the dynamic streaming model.

\end{abstract}
\end{titlepage}

    \include*{intro}
    \include*{prelims}

    \include*{general}
    \include*{emd}

    \include*{median}
    \include*{appls}
    \include*{acks}

    \bibliographystyle{alpha}
    \bibliography{ref}

    \appendix

\include*{appendix}

\end{document}

%% file: intro.tex
\ir{TODO: sketches vs embeddings (NO NEED TO DO THIS), open problem from FS (DONE), pad tables a bit, state the running time (DONE BUT THE TEXT IN SECTIONS 3.1 and D SOMEWHAT SKETCHY), emphasize the case of small $k$ (DONE)}

\section{Introduction}

The field of sparse recovery studies the following question: for a signal $x$, when is it possible to  compute an approximation $\hat{x}$ to $x$ that is parameterized by only a small number coefficients, given only a small number of linear measurements of $x$? The answers to this basic question, i.e., the {\em sparse recovery schemes}, have found a surprising number of applications in a broad spectrum of fields, including  {\em compressive sensing} ~\cite{CRT06:Stable-Signal, Don06:Compressed-Sensing},  data stream computing~\cite{muthukrishnan2005data} (see also the resources at {\tt sublinear.info}) and Fourier sampling~\cite{gilbert2014recent}.

A particularly useful and well-studied formalization of this question is that of {\em stable sparse recovery}.  A general formulation of the problem is as follows.  For a norm $\|\cdot\|$, sparsity parameter $k$, probability of failure $P$ and an approximation factor $K>0$, design a distribution over $m \times n$ matrices $A$ which has the following property: 
\begin{quote}
There is an algorithm $\mathcal{A}$ that, for any $x$, given $Ax$,  recovers a vector $\widehat{x}=\mathcal{A}(Ax)$  such that
\begin{equation}
\label{e:approx}
\|x-\widehat{x}\| \le K \cdot \min_{k\text{-sparse } x'}  \|x-x'\|
\end{equation}
with probability at least $1-P$. 
\end{quote}
Here we say that $x'$ is $k$-sparse if it has at most $k$ non-zero coordinates\footnote{Further generalizations of the problem can be obtained by allowing the sparsity in arbitrary basis, or by allowing different norms on the LHS and RHS of Equation~\ref{e:approx}. Although important, we will not consider these generalizations in this paper.}. The typical choices of the norm $\|\cdot\|$ are either $\ell_1$ or $\ell_2$. However, several other variants have been studied as well: \cite{BGIKS,allen2014restricted} studied sparse recovery under general $\ell_p$ norms, \cite{gupta2010sparse,indyk2011k,mo2013compressive} considered the Earth-Mover-Distance (EMD) norm, while \cite{ward2014unified} considered rearrangement-invariant block norms.

It is easy to observe that the number of measurements $m$  must depend on the sparsity parameter~$k$: the more information about the signal we want to acquire, the more measurements must be taken. For $\ell_1$ and $\ell_2$ norms, the tradeoff between $m$ and $k$ is well-understood: it is known that $m=O(k\log (n/k))$ measurements suffice~\cite{CRT06:Stable-Signal}, and  this bound is tight~\cite{Don06:Compressed-Sensing,do2010lower}. For other norms, however, our understanding of the tradeoffs is much more limited. 

\subsection{Our results}
 In this paper we initiate the study of sparsity-measurements trade-offs for \emph{generic} norms\footnote{In fact, our results hold even for {\em quasi-norms}, e.g., $\ell_p$ norms for $p<1$ (see Preliminaries for more details). However,  for the sake of simplicity, in the rest of the paper we will mostly focus on norms.}.  Our results generalize the previously known tradeoffs, and provide improved bounds for specific norms, notably EMD and $\ell_p$ for $p  \in (0,1)$. Further, our results for EMD immediately yield new sketching algorithms and new lower bounds for the low-dimensional $k$-median clustering problem. 

Our first result shows that, for norms that \emph{admit efficient linear sketches} the number of measurements sufficient for sparse recovery is closely related to the doubling dimension of $k$-sparse vectors under that norm.  Formally, we prove the following theorem.

\begin{theorem}
\label{t:upper}
    Suppose that $X = (\Rbb^n, \|\cdot\|)$ is an $n$-dimensional normed space and $1 \leq k \leq n$ be the sparsity parameter. 
    Assume that, for some (distortion) parameter $D \ge 1$  there is a distribution over $s \times n$ random  (sketch) matrices $S$ and an (estimator) function $E: \Rbb^s \to \Rbb$ such that 
for any   $x$ and a $k$-sparse $y$ we have
\[ \Pr[\|x-y\| \le  E(Sx,Sy) \le D \|x-y\| ]\ge 2/3 .\]
    Furthermore, let $d$ be the doubling dimension of the set of $k$-sparse vectors
    from $\Rbb^n$ with respect to the metric induced by $\|\cdot\|$.
    Then, for every $0 < \eps, \tau < 1/3$ 
    there exists a distribution over random matrices $A \in \Rbb^{m \times n}$
    with
    $$
        m = O\Big(s \cdot \Big(d \cdot \log (D / \eps) +
        \log \log (1 / \tau)\Big) \Big)
    $$
    such that for every $x \in \Rbb^n$
    given $Ax$ we can recover with probability at least $2/3$
    a vector $\widehat{x} \in \Rbb^n$
    such that
    \begin{eqnarray}
\label{e:approx2}
        \|x - \widehat{x}\| \leq  
        (1 + \eps) D \min_{\mbox{$k$-sparse $x^*$}} \|x - x^*\| + \tau \|x\|.
    \end{eqnarray}
\end{theorem}

To explain the theorem, we first observe that the guarantee given by~(\ref{e:approx2}) is analogous to the one given by~(\ref{e:approx}), with the exception of the extra additive term $\tau \|x\|$. The ``precision parameter'' $\tau$ can be made arbitrarily small, at a price  of increasing the number of measurements by an extra $ \log \log (1 / \tau)$ term. Similar tradeoffs between the precision and the number of measurements are quite common in compressive sensing schemes\footnote{E.g., in most of the existing sparse Fourier transform algorithms the sample complexity depends logarithmically on the precision parameter~\cite{gilbert2014recent}.}, although we do not know whether this extra term is necessary in our setting.  Apart from the precision dependence, the number of measurements $m$ is linear in the doubling dimension $d$, linear in the sketch length $s$ and logarithmic in the distortion~$D$. 

Our theorem requires that the normed space of interest admits efficient linear sketches.  We believe that some variant of this assumption is necessary for sparse recovery, as such sketches are needed if one wants to estimate the approximation error, i.e., the RHS of Equation~\ref{e:approx2}. However, this intuition does not lend itself to a formal argument, as  e.g., for the $\ell_1$ norm there exist  sparse recovery schemes~\cite{CRT06:Stable-Signal, Don06:Compressed-Sensing} that satisfy Equation~\ref{e:approx2} {\em without} explicitly estimating the approximation error. Still, the $\ell_1$ norm supports efficient sketches, which suggests that some form of sketchability of the norm could be a necessary condition.

To illustrate Theorem~\ref{t:upper}, consider the case of the $\ell_p$ (quasi)-norms for $p \in [0,2]$. It is known~\cite{indyksketch} that these norms allow sketches with distortion $D=1+\epsilon$ and dimension $s=O(1/\epsilon^2)$ for any $\epsilon>0$, and it is also immediate that the doubling dimension $d$ is 
$O(k \log(n/k))$. Therefore, for $p \in [1,2]$ our theorem reproduces the known optimal $O(k \log (n/k))$ measurement bound, up to the dependence on the precision parameter $\tau$.
The same bound is obtained for $p \in (0,1)$. The latter result is, to the best of our knowledge, new.

We note that Theorem~\ref{t:upper} is not efficient: it does not provide a \emph{polynomial time} algorithm for recovering $\widehat{x}$ from $Ax$. Given the generality of the setting, in particular, the fact that it allows a general (sketchable) norm $\|\cdot\|$, we believe that a general polynomial time recovery algorithm is unlikely to exist. However, it is possible that efficient algorithms exist for specific norms which have good computational properties. For example, we show that for the case of the Earth-Mover Distance norm discussed in more detail below, the recovery algorithm runs in time polynomial in $n$ and $\log^k n$. In particular, the running time is polynomial for any constant $k$.

\paragraph{Lower bound} The $\ell_p$ norm example shows that the bound of Theorem~\ref{t:upper} is tight for {\em some} norms. In fact, one can show that the linear dependence on the doubling dimension $d$ is necessary for {\em all} norms whose ``aspect ratio'' is bounded by a polynomial in $n$. In particular, we show the following theorem.

\begin{theorem}
\label{t:lower}
  Consider any norm $\norm{\cdot}$ over $\R^n$ for which
  $\frac{1}{n^c} \leq \frac{\norm{x}}{\norm{x}_2} \leq n^c$ for some
  constant $c$.  Let $T_k \subset \R^n$ denote the set of $k$-sparse
  vectors and $d > 1$ denote the doubling dimension of $[0, \infty)^n
  \cap T_k$ with $\norm{\cdot}$.  Then any sparse recovery scheme 
  for $[0,\infty)^n$ with approximation factor $K$ requires 
  $m =\Omega(d / \log K)$ measurements.
\end{theorem}

Note that the theorem holds even for vectors $x \ge 0$, which will be useful in the context of the Earth-Mover Distance.

\paragraph{Earth-Mover Distance}  Our results have direct implications for sparse recovery over the Earth-Mover Distance (EMD) norm. This norm is defined over $n$-dimensional vectors with $n=\Delta^2$, where such vectors can be interpreted as functions $[\Delta]^2 \to \R$.  Informally, for vectors $x,y :[\Delta]^2 \to \R_+$ which have the same $\ell_1$ norm, the EMD is
defined as the cost of the min-cost flow that transforms $x$ into $y$,
where the cost of transporting a ``unit'' of mass from a point $p \in
[\Delta]^2 $ of $x$ to a point $q \in [\Delta]^2$ of $y$ is equal to
the $\ell_1$ distance\footnote{One can also use the $\ell_2$ distance.
 Note that the two distances
  differ by at most a factor of $\sqrt{2}$ for two-dimensional
  images.}  between $p$ and $q$. See Preliminaries for a formal
definition. 

Earth-Mover Distance and its variants are popular metrics for
estimating similarity between images and feature
sets~\cite{RTG,GD}. Furthermore, the $k$-sparse approximation of
non-negative vectors under the EMD norm has the following natural
interpretation. Let $\hat{x}$ be the $k$-sparse vector closest to $x$
under this norm.
Then one can observe that the
non-zero entries of $\hat{x}$ correspond to the cluster centers in the best
$k$-median\footnote{For completeness, in our context the $k$-median clustering problem is
  defined as follows. First, each pixel $p \in [\Delta]^2$ is interpreted
  as a point with weight $x_p$.
Then the goal is to find a set $C \subset [\Delta]^2$ of $k$ ``medians'' that
minimizes the objective function $\sum_{p \in [\Delta]^2} x_p \cdot \min_{c \in C}
\|p-c\|_2$.} clustering of $x$.  
 Thus, sparse recovery schemes for the EMD norm provide methods for recovering near-optimal solutions to the planar $k$-median problem from few linear measurements of the input point-sets, a problem that has attracted a considerable attention in streaming and sketching literature~\cite{I04,FSoh,indyk2011k}. 

The state of the art schemes for this problem are listed in Figure~\ref{f:results}. In particular, the best known bound for the number of measurements is $O(k \log (n/k) )$, which mimics the best possible bound achievable for sparse recovery in the $\ell_1$ norm.

\begin{figure}
\begin{center}
\begin{tabular}{|c|c|c|c|}
\hline
Randomized/Deterministic	& Sketch length $m$ & 	Approximation factor \\
 \hline
 Deterministic & $k \log n \log(n/k)$ & $O(1)$\\
 Deterministic & $k  \log(n/k)$  & $\sqrt{\log (n/k)}$\\
 Randomized  & $k \log(n/k)$ & $O(1)$\\
 \hline
 \end{tabular}
 \caption{Performance of sparse recovery schemes for the EMD from \cite{indyk2011k}. The schemes assume that the input vector $x$ is non-negative. Each result implies a sketching scheme for the $k$-median problem with the same parameters.}
 \label{f:results}
 \end{center}
 \end{figure}

We show that Theorem~\ref{t:upper}  provides new results for this problem. Specifically, we show that the doubling dimension of the EMD norm over $k$-sparse vectors is only $O(k \log \log n)$. 
Combined with the known fact that the EMD norm can be embedded into $\ell_1$ with distortion $O(\log n)$ \cite{c-setfr-02, it-fire-03} (and therefore its sketching complexity $s$ is constant),  this implies that there exist a  sparse recovery scheme for EMD with approximation factor $O(\log n)$ that uses only $O(k (\log \log n)^2)$ measurements (ignoring the dependence on the precision). The running time of recovery procedure is polynomial in $\Delta$ and $\log^k \Delta$ (again ignoring the dependence on the precision), which is polynomial in $\Delta$ for any $k$ up to $\log \Delta/\log \log \Delta$. We further show that the result can be strengthened in three ways:
\begin{itemize}
\item By performing a more careful analysis of the embedding procedure of~\cite{it-fire-03}, we show that it in fact incurs a distortion of $O(\log k + \log \log n)$ with constant probability, which is sufficient for our purposes.
\item By using a variant of the embedding (given in~\cite{i-nltcf-07}) and combining it with a sketch of~\cite{vz-rsdre-12}, we show the distortion can be reduced further to $O(\log k)$ while increasing the sketch length by a factor of $O(\log^{\delta} n)$  for any constant  $\delta>0$. Note that in the case of \emph{constant}~$k$, the approximation we obtain is constant as well.
\item Finally, we consider vectors $x$ with the property that, for some integer $N$, all entries $x_p$ are multiples of $1/N$ (in this case we say that $x$ has {\em granularity} $1/N$). Such vectors correspond to characteristic vectors of multisets of size $N$, and naturally occur in the \emph{unweighted} $k$-median problem over point sets of size $N$. In this case we show that, in the bounds for the doubling dimension and distortion, we can replace $\log \log n$ by $\log \log N$. Notably, the bounds we obtain in this case \emph{are independent of the ambient dimension $n$}.
\end{itemize}

By combining these bounds with Theorem~\ref{t:upper} we obtain sparse recovery schemes for EMD with the guarantees as in Figure~\ref{f:ourresults} (see also Section~\ref{s:sremd} for the formal statement of the results).

\begin{figure}[h]
\begin{center}
\begin{tabular}{|c|c|c|c|}
\hline
Randomized/Deterministic	& Sketch length $m$ & 	Approximation factor \\
 \hline
 Randomized  & $k (\log \log n)(\log (\log k + \log \log n))+  \log \log (1 / \tau)$ & $O(\log k + \log \log n)$\\
 Randomized & $k \log^{\delta} n +  \log \log (1 / \tau)$ & $O(\log k)$\\ 
Randomized, lower bound & $\Omega(k(\log \log (n/k))/\log K)$ & $K\geq 2$\\
 \hline
 \end{tabular}
 \caption{Performance of our sparse recovery schemes for the EMD.  The schemes assume that the input vector $x$ is non-negative. 
The first two results imply a sketching scheme for the $k$-median problem with the same parameters.}
 \label{f:ourresults}
 \end{center}
 \end{figure}

The aforementioned bounds are quite surprising, as they are provably impossible to achieve for the $\ell_1$ or $\ell_2$ norms. In particular---for $\ell_1$ and $\ell_2$---one needs $\Omega(\log n)$ measurements to achieve constant approximation factor even for $k=1$, and $\Omega(\log n/\log \log n)$ measurements to achieve $O(\log n)$ distortion~\cite{do2010lower}. This means that the EMD norm is actually {\em easier} than $\ell_p$ norms from the sparse recovery perspective, at least in a range of parameters. 

We also show that at least one $\log \log n$ factor in the measurement bound is necessary as long as $k\ge 2$, by proving a lower bound for the doubling dimension of $k$-sparse vectors under EMD and using Theorem~\ref{t:lower}. In fact, our lower bound argument applies almost verbatim to the space complexity of the following {\em data stream} problem: design a data structure that maintains a vector $x$ under increments and decrements of its coordinates which, when queried, reports a $k$-sparse approximation to $x$ with approximation factor $K$ with a constant probability. As discussed earlier, in the context of the EMD norm this task corresponds to the problem of maintaining a $k$-median clustering of a dynamic point set where points can be inserted and deleted (i.e., the coordinates of $x$ can be incremented and decremented). As we show in Theorem~\ref{streaming_lb}, the space \emph{bit} complexity of this problem is $\Omega(\frac{d}{\log K} \log n)$ for general norms, thus, in particular, $\Omega(\frac{k}{\log K} \log(\log(\frac{\Delta^2}{k})) \cdot \log \Delta)$ for the EMD norm. The last bound addresses the open question of~\cite{FSoh} (Section 7)  who asked whether  it is possible to maintain a constant size (for fixed $k$ and $K$) ``core-set''\footnote{Informally, a core-set for the $k$-median problem over a set of  points $P$ is a weighted subset $C \subset P$ such that a solution to $C$ provides an approximate solution to $P$. Core-sets provide a tool for solving streaming problem  for $k$-means and $k$-median problems in data streams. See~\cite{FSoh} for more details.}  for the $k$-median and $k$-means problem in dynamic data streams. 
Although our  arguments do not consider the core-set size per se, we do show that any algorithm that solves $k$-median and $k$-means \footnote{The lower bound for the $k$-means problem is presented in Appendix~\ref{s:kmeans}.}  in the dynamic data stream model must use a {\em super-constant} number of words of size $\log \Delta$, even for constant $k$ and $K$.

Finally, we show that for the case of $k=1$, a sparse recovery scheme exists with $O(1)$ measurements for constant $d$ and $\eps$, independent of $n$. This is again in sharp contrast to $\ell_1$ or $\ell_2$ norms, as well as the aforementioned case of $k \ge 2$.
 
\subsection{Our techniques and related work} 

Our upper bound for the number of measurements relies on the connection between sparse recovery and the approximate nearest neighbor search.  Specifically, our goal can be phrased as finding the nearest neighbor of $x$ in a set of bounded doubling dimension. The latter problem can be solved using the {\em navigating nets} data structure~\cite{kl-nnsap-04}, and indeed we are using a similar top-down search approach in our algorithm. However, we need to deal with complications that arise due to the fact that in our setting we can only estimate distances approximately and with a certain probability. Specifically, to obtain the desired bound, we need to ensure that the total number of distances that our sketch needs to preserve is only {\em linear} in the depth of the tree. This allows us to bound the probability of failure of the algorithm by taking the union bound over a small number of events. It is easy to observe, however, that the path in the tree taken by the search algorithm is {\em adaptive}, i.e., the approximation errors incurred by the sketch at one level affect the points considered by the algorithm at the next level.  Nevertheless we show that the path cannot be \emph{too adaptive}, and that one can identify a set of points of size linear in the tree depth so that preserving all the distances from those points  to $x$ ensures the correctness of the algorithm.  The details are in Section~\ref{s:upper}

Our lower bound builds on the argument from~\cite{do2010lower}, where the number of measurements was lower bounded by encoding long bit sequences into the signal $x$, such that those bits could be unambiguously decoded by the sparse recovery algorithm. The encoding proceeded on several distance scales. At each scale, the encoding used a large set of almost equidistant $k$-sparse vectors as the ``dictionary''. Since the maximum size of such sets is directly related to the doubling dimension of the space, the lower bound argument goes through in the setting of a
\emph{general} norm. The details are in Section~\ref{s:lower}.

The doubling dimension of the set of $k$-sparse vectors under EMD was
previously studied by~\cite{gkk-ecmd-10}, who showed that it is at
most $O(k \log k)$ for the special case of measures induced by $k$-sets, i.e., measures of granularity $1/k$.
For this case it is in fact not difficult to improve the
bound to $O(k)$ and we give an outline of the improved argument in
Section~\ref{s:emd}. However, for our applications we need a bound
that holds for general measures. This makes the argument more
complex, since we need to deal with general flows. In both cases the
idea of the proof is to explicitly construct a covering of a ball of
radius $R$ using a small number of balls of radius $R/2$, by using the
geometric and combinatorial properties of planar flows. The details are in Section~\ref{s:emd} (for general non-negative vectors) and Section~\ref{s:additional} (for vectors of bounded granularity).

Our improved analysis of the embedding of~\cite{it-fire-03}, as well as the analysis of the embedding from~\cite{i-nltcf-07}, utilize the fact that our application allows us to relax the standard embedding definition in two ways. First, we only need to preserve the distance between a $k$-sparse vector and a general vector, as opposed to between any pair of vectors (see the statement of Theorem~\ref{t:upper} for the precise guarantee that we are after). Second, we only need to ensure that the distances are preserved with constant probability, not in expectation, which means that we can tolerate events that incur high distortion as long as they occur with low enough probability. Combining the two relaxations\footnote{It can be seen that both relaxations are needed in order to achieve the better bound. In particular, the {\em expected} distortion of the embedding is $\Theta(\log n)$, even for a pair of $1$-sparse vectors. 
Similarly, if $k=n$, the distortion of the embedding is $\Omega(\log n)$ with probability $1-o(1)$.} with a more careful analysis allows us to achieve the improved bound, surprisingly almost without any modifications to the embeddings themselves. The details are in Section~\ref{s:sketches}.

We note that if one wants to preserve EMD between two vectors that are {\em both} $k$-sparse, then one can embed those vectors into $\ell_{1-\epsilon}$ with distortion $O(\log k)$~\cite{bavckurs2014better}, which yields a sketch with the same distortion and constant size~\cite{indyksketch}. Also, for the case when one of the vectors is $k$-sparse, a recent work~\cite{yousefi2014improved} shows a sketch with distortion $O(\min(k^3,\log n))$ and size roughly $O(\log^4 n)$. The sketch in this paper substantially improves over the latter bound.

For the $1$-median problem we solve an $\ell_1$-regression problem. We give oblivious sketches that provide subspace embeddings for the $\ell_1$-norm for $d$-dimensional subspaces with a ``disjoint basis'' property that arises in this setting. Our embedding works when the basis is expressible as the union of a small number of sets of vectors, where in each set the vectors have disjoint support. Unlike existing oblivious embeddings for $\ell_1$ \cite{cdmmmw13,mm13,sw11,wz13}, we obtain $(1+\eps)$ instead of $\poly(d)$ distortion, and low $\delta$ instead of constant probability of failure (to simultaneously preserve norms of all vectors in the space). Our embedding maps $n$-dimensional vectors to $O(d/\eps^2 \log(d/(\delta \eps)))$ dimensions. We overcome non-embeddability results for $\ell_1$ \cite{bc05,cs02} by using a non-convex estimator. This is reminiscent of estimators for data streams \cite{indyksketch}, but complicated here by the 
fact that we require the stronger notion of a subspace embedding. 
It is known (see, e.g., \cite{abs10}) that for constant $d$ and $\eps$ one can solve the $1$-median by taking $O(1)$ samples and solving the problem on the samples, but this cannot be expressed as a linear sketch with fewer than $\Omega(\log n)$ measurements (the sampling lower bound follows from Theorem 8 of \cite{jst11}), whereas we achieve $O(1)$ measurements. 
The details are in Section~\ref{s:1median}.

%% file: prelims.tex
\section{Preliminaries}

\paragraph{EMD.} We start by defining EMD. Consider any two
non-negative vectors $x,y : [\Delta]^d \to \R_+$ such that
$\|x\|_1=\|y\|_1$.  Let $\Gamma(x,y)$ be a set of functions
$\gamma: [\Delta]^d \times [\Delta]^d \to \R_+$, such that for any
$i,j \in [\Delta]^d$ we have $\sum_l \gamma(i,l)=x_i$ and $\sum_l
\gamma(l,j)=y_{j}$.  Then we define
\[ \EMD^*(x,y) = \inf_{\gamma} \sum_{i,j \in [\Delta]^d } \gamma(i,j)\|i-j\|_1 \]
Note that if $x$ and $y$ are characteristic vectors of some sets $A, B
\subset[\Delta]^d$, then $\EMD^*(x,y)$ is equal to the value of the
minimum cost matching between $A$ and $B$.

For the case of general vectors $x,y$, we define
\[ \EMD(x,y) = \inf_{\substack{x' \le x, y' \le y\\ \|x'\|_1=\|y'\|_1}} \EMD^*(x',y')
+ D[ \|x-x'\|_1 + \|y-y'\|_1 ] \]
where $D=d\Delta$ is the diameter of the set $[\Delta]^d$.

\paragraph{Metric spaces.} For a metric space $(X,d_X)$, we define $B_X(u,r)$ or, equivalently, $\Ball_X(u,r)$ to be the ball  centered at $u$ or radius $r$ containing all points from $X$ within $r$ from $u$: $B_X(u,r):=\{x \in X:d_X(u,x)\leq r\}$. Further, 
for a metric space $(X,d_X)$, the {\em doubling dimension} is the smallest number $d$ such that, for every
$r>0$ and any $x \in X$, we can choose $x_1, x_2, ..., x_{2^d}\in X$ with
$$B_X(x,r)\subseteq B_X(x_1,r/2)\cup B_X(x_2,r/2) \cup ... \cup B_X(x_{2^d},r/2).$$

Finally, for $K \geq 1$ we define a \emph{$K$-quasi-metric space} as a variant of a metric space, where we have the following
\emph{relaxed} triangle inequality:
$
    d(x, y) \leq K \cdot \bigl(d(x, z) + d(z, y)\bigr).
$
Thus, every metric space is a $1$-quasi-metric space. We define $K$-quasi-norms in an analogous way.

%% file: general.tex
\section{Upper Bound on Measurement Complexity}
\label{s:upper}

Suppose we have a $K$-quasi-metric space $\Mc = (X, \rho)$ \ir{Define $K$-metric spaces!}
and a closed subset $Y \subseteq X$ with doubling dimension $d$.
Let us assume we can sketch distances
between points from $X$ and $Y$
with distortion $D$, sketch size $s$ and success probability at least $2 / 3$ (see Theorem~\ref{t:upper} for the formal definition).

The following Lemma builds on a result from~\cite{kl-nnsap-04} on approximate nearest neighbor search in doubling spaces.

\begin{lemma}
    \label{recovery_upper}
    For every $0 < \eps < 1 / 2$, $0 < \lambda < \Lambda$ and $y_0 \in Y$
    one can sketch points of $X$ with sketch size
    $$
        O\Big(s \cdot \Big(d \log (DK / \eps) +
        \log \log (\Lambda / \lambda)\Big)\Big)
    $$
    so that from this sketch for $x \in X$ with $\rho(x, y_0) \leq \Lambda$ we can recover
    with probability at least $2/3$
    a point $\hat{y} \in Y$ such that
    \begin{equation}
        \label{recovery_guarantee}
        \rho(x, \hat{y}) \leq
        \max((1 + \eps) DK \cdot \rho(x, Y), \lambda).
    \end{equation}
\end{lemma}

\begin{proof}
First, we describe the recovery procedure and then show how to sketch
points.
For now, we assume that for the point of interest $x \in X$ and for every $y \in Y$ we know a number
$q(y)$ such that
\begin{equation}
    \label{distortion}
    \rho(x, y) \leq q(y) \leq D \cdot \rho(x, y).
\end{equation}

The recovery procedure we describe has several parameters: a positive integer $L$, a real $0 < \alpha < 1$ and real $\beta, \gamma > D$.
For the reasons that will be clear later we require that
\begin{equation}
    \label{technical_condition}
    K^2 \cdot (\alpha + 2 \gamma) \leq \alpha \beta.
\end{equation}

The recovery procedure is as follows. First, for every $0 \leq i \leq L$
we build a $r_i$-net $N_i$ of $Y \cap B_X(y_0, \Lambda)$, where $r_i = 2 \alpha^i \Lambda$
such that all pairs of points from $N_i$ have pairwise distances larger than
$r_i$. In particular, $|N_0| = 1$, and for every $i$ the size of $N_i$ is finite since the doubling dimension $d$ of $Y$ is finite.
Such a net can be found using a straightforward greedy algorithm.
Second, given a point $x \in X$ with $\rho(x, y_0) \leq \Lambda$ we
recover an approximate nearest neighbor from $Y$ as follows:
\begin{figure}
\begin{algorithmic}
   \State Search Procedure:
    \State $y_0 \gets \mbox{the only element of $N_0$}$
    \For{$i \gets 1 \ldots L$}
        \State $S_i \gets N_i \cap B_X(y_{i - 1}, \beta r_i)$
        \State $y_i \gets
        \argmin_{y \in S_i}
        q(y)$
        \If{$q(y_i) > \gamma r_i$}
            \State\Return{$y_{i - 1}$}
        \EndIf
    \EndFor
    \State\Return{$y_L$}
\end{algorithmic}
\end{figure}

Now let us analyze this procedure. Denote $y^* = \argmin_{y \in Y}
\rho(x, y)$ one of the nearest neighbors for $x$ from $Y$ (note that $y^*$ exists, since $Y$ is assumed to
be closed). The proof follows from the following three claims (the proofs are in Appendix~\ref{s:proofs}).

\begin{claim}
    \label{key_estimate}
    If~(\ref{technical_condition})
    holds and for some $1 \leq i \leq L$ one has
    $q(y_{i - 1}) \leq \gamma r_{i - 1}$, then
    $\rho(y^*, S_i) \leq r_i$.
\end{claim}

Now let us analyze the case when the algorithm returns $y_{i - 1}$
for some $1 \leq i \leq L$.

\begin{claim}
    \label{return_yi}
    If~(\ref{technical_condition}) holds and
    the algorithm returns $y_{i - 1}$ for some
    $1 \leq i \leq L$, then
    $$
        \frac{\rho(x, y_{i - 1})}{\rho(x, y^*)} < \frac{D K \gamma}
        {\alpha (\gamma - DK)}.
    $$
\end{claim}

Next, suppose that our algorithm returns $y_L$.

\begin{claim}
    \label{return_yl}
    If~(\ref{technical_condition}) holds and the algorithm returns
    $y_L$, then
    $
        \rho(x, y_L) \leq 2 \gamma \alpha^L \Lambda.
    $
\end{claim}

Let us now show how to set $L$, $\alpha$, $\beta$ and $\gamma$.
Claims~\ref{return_yi} and~\ref{return_yl} imply that in order to
satisfy~(\ref{recovery_guarantee}) we need to satisfy together
with~(\ref{technical_condition}) the following conditions:
\begin{eqnarray}
    \label{eps_condition}
    \frac{D K \gamma}{\alpha (\gamma - D K)} &\leq& (1 + \eps) D K, \\
    \label{delta_condition}
    2 \gamma \alpha^L \Lambda &\leq& \lambda.
\end{eqnarray}
It is immediate to see that we can satisfy~(\ref{technical_condition}),
(\ref{eps_condition}) and~(\ref{delta_condition})
simultaneously by setting
$\alpha = 1 - \Theta(\eps)$, $\beta = \Theta(DK^3 / \eps)$,
$\gamma = \Theta(DK / \eps)$ and
$
    L = \Theta\Big(\frac{1}{\eps} \cdot
    \log \frac{D K \Lambda}{\eps \lambda}\Big).
$

So far we assumed that we have access to a function $q(\cdot)$
that satisfies~(\ref{distortion}).
In reality we build such a function from sketches of distances
between points from $X$ and $Y$.
Suppose we can build a subset\footnote{Note that $Q$ is more than just a single path from the ``root'' to the solution, as the behavior of the algorithm is not deterministic and depends on the random bits chosen by the sketching procedure.} $Q \subseteq Y$ with $|Q| \leq N$ such
that for a given $x \in X$ the recovery procedure can
query $q(y)$ only for $y \in Q$.
Then, we can use the standard amplification argument for the median
estimator, and sketch $x$ in size $O(s \log N)$ to get
a randomized function $q'(\cdot)$ such that for every $y \in Q$ one has
$
    \Prb{}{\rho(x, y) \leq q'(y) \leq D \cdot \rho(x, y)} \geq
    1 - \frac{1}{3N}.
$
Now we use $q'(\cdot)$ for the recovery and by the union bound the recovery algorithm
succeeds with probability at least $2/3$.
It is only left to upper bound $N$ for an appropriately chosen set $Q$.

It is clear that we query $q(\cdot)$ for points only from
$$
    \bigcup_{\begin{smallmatrix}i \in [L] \\
    q(x, y_{i - 1}) \leq \gamma \cdot r_{i - 1}\end{smallmatrix}} S_i
    \subseteq
    \bigcup_{\begin{smallmatrix}i \in [L] \\
    \rho(x, y_{i - 1}) \leq \gamma \cdot r_{i - 1}\end{smallmatrix}} S_i
$$
(the inclusion is by~(\ref{distortion})).
By Claim~\ref{key_estimate} the right-hand side is included in
$$
    Q = \bigcup_{1 \leq i \leq L}
        \big(N_i \cap B_X(y^*, K^2 \cdot (1 + 2 \beta) \cdot r_i)\big).
$$
Since $Y$ has doubling dimension $d$ and points from $N_i$ are $r_i$-separated, we get
$N = |Q| \leq L \cdot \bigl(K^2 \cdot (1 + 2 \beta)\bigr)^{O(d)}$. Now using the values of
$L$, $\alpha$, $\beta$ and $\gamma$, we get that the final sketch size is
$
    O(s \log N) \leq O(s \cdot (d \log (K \cdot (1 + \beta)) + \log L))
    \leq O\Big(s \cdot \Big(d \log (DK / \eps) +
    \log \log (\Lambda / \lambda)\Big) \Big).
$
\end{proof}

\begin{corollary}
Suppose that $X$ is induced by a norm of dimension $n$, and that there is an algorithm that computes the sets $S_i$ defined by the search procedure in time $|S_i|^{O(1)}$. 
Then the search procedure runs in time polynomial in $N =L \cdot \bigl(K^2 \cdot (1 + 2 \beta)\bigr)^{O(d)}$ and $n$.
\end{corollary}



%% file: emd.tex
\subsection{Upper bound on the doubling dimension of $\EMD$}
\label{s:emd}

We will prove that the doubling dimension of
$k$-sparse probability measures over $[\Delta]^2$ equipped with $\EMD$
is $O(k \log \log \Delta)$. For a weaker and simple bound $O(k \log k)$ on
the doubling dimension in the case of $k$-sparse \emph{subsets},
see \cite{gkk-ecmd-10}. In fact, it is not hard to prove upper bound $O(k)$ on
the doubling dimension for $k$-sparse subsets. Notice that in this case
the upper bound on the doubling dimension does not depend on the size of 
the grid. We will now provide an intuition why the upper bound $O(k)$ holds.

We have an $\EMD$ ball $\Ball_{\EMD}(\mu, R)$ or radius $R$ centered at 
$k$-sparse measure $\mu$ such that
$\mu(x,y)=1$ for all $(x,y)\in \supp(\mu)$. (We can think of $\mu$ as a $k$-sparse
set.) And we would like to cover all $k$-sparse subsets within 
$\Ball_{\EMD}(\mu, R)$ with $2^{O(k)}$
$\EMD$ balls of radius $R/2$ centered at $k$-sparse subsets.

First, let's show how to cover all subsets $o \in \Ball_{\EMD}(\mu,R)$
with $o$ satisfying $\|o_i-\mu_i\|_1=\Theta(R/k)$ for all $i \in [k]$.
$o_i$ and $\mu_i$ denote points in $\supp(o)$ and $\supp(\mu_i)$ and they
get matched togefther in the optimal transportation between $o$ and $\mu$.
For this, we take $R/(100k)$-net of $\Ball_{\ell_1}(\mu_i,10R/k)$
for every $i \in [k]$. Every such net is of size $O(1)$. To
cover all the $o \in \Ball_{\EMD}(\mu,R)$, we need to take a
representative from a net from $\Ball_{\ell_1}(\mu_i,10R/k)$ for all $i \in [k]$
and combine representatives in $k$-sparse subsets. There are $2^{O(k)}$ possible
ways to construct subsets by taking representatives.

In the case when we do not have the mentioned guarantee at the beginning of
the previous paragraph, we can guess values $\|o_i-\mu_i\|_1$ up to a constant
factor and construct covers for all guesses. We need to show that 
it is enough to take at most $2^{O(k)}$ guesses. And it can
indeed be shown by noticing that we do not need to cover $\ell_1$ balls
of very small radius (when $\|o_i-\mu_i\|_1$ is small).

We proceed by showing upper bound on the doubling dimension when
we consider arbitrary measures with support of size at most $k$ living
in a square of side length $\Delta$.

\begin{lemma}
	\label{doubling_grid}
    The doubling dimension of the set of $k$-sparse probability
    measures over
    $[\Delta]^2$ under EMD metric is $O(k \log \log \Delta)$.
\end{lemma}
\begin{proof}
Let $\mu$ be a $k$-sparse probability measure over $[\Delta]^2$ and let $R > 0$ be some real number.
Our goal is to cover $B_{\EMD}(\mu, R)$ with $\log^{O(k)} \Delta$ EMD-balls centered in $k$-sparse measures
and of radius $R / 2$.
In order to achieve this it is sufficient to cover $B_{\EMD}(\mu, R)$ with $\log^{O(k)} \Delta$ EMD-balls
centered in \emph{arbitrary} measures and of radius $R / 4$.

The pseudocode in Figure~\ref{f:pseudo} builds a set of measures $\Mc$
that serve as centers of balls with radius $R / 4$ that together cover
$B_{\EMD}(\mu, R)$. Roughly speaking, we first guess the topology of the optimal flow.
Then we guess the lengths of the corresponding edges. Then we
guess the support. And finally we guess the masses transported
over the edges.

We assume that {\sc BuildNet}$(p, r)$ returns an $(r / 100)$-net of $B_{\ell_1^2}(p, r) \cap [\Delta]^2$.
\begin{figure}[h]
\begin{algorithmic}[1]
    \State $m_0 \gets R / (100 \Delta k)$
    \State $\Mc \gets \emptyset$
    \For{$c \colon \supp \mu \to \Zbb_{>0}$ such that $\sum_{(x, y) \in \supp \mu} c(x, y) \leq 2k$}
        \State $\Ic \gets \set{(x, y, i) \mid (x, y) \in \supp \mu, 1 \leq i \leq c(x, y)}$
        \For{$l \colon \Ic \to \set{1, 1.01, 1.01^2, \ldots, 2 \Delta}$}
            \For{ $(x, y, i) \in \Ic$ and for all $p(x, y, i) \in \mbox{\Call{BuildNet}{$(x, y)$, $l(x, y, i)$}}$}
                \For{$m \colon \Ic \to \set{0, m_0, 1.01 \cdot m_0, 1.01^2 \cdot m_0, \ldots, \min(1, R)}$}
                    \If{for every $(x, y) \in \supp \mu$ we have $\sum_{i \colon (x, y, i) \in \Ic} m(x, y, i) \leq \mu(x, y)$}
                        \State let $\mu'$ be a measure over $[\Delta]^2$ that is identically zero
                        \For{$(x, y) \in \supp \mu$}
                            \State $s \gets 0$
                            \For{$i : (x, y, i) \in \Ic$}
                                \State $s \gets s + m(x, y, i)$
                                \State $\mu'(p(x, y, i)) \gets \mu'(p(x, y, i)) + m(x, y, i)$
                            \EndFor
                            \State $\mu'(x, y) \gets \mu'(x, y) + \mu(x, y) - s$
                        \EndFor
                        \State $\Mc \gets \Mc \cup \set{\mu'}$
                    \EndIf
                \EndFor
            \EndFor
        \EndFor
    \EndFor
\end{algorithmic}
\caption{Pseudocode for net construction}
\label{f:pseudo}
\end{figure}
It is immediate that $|\Mc| \leq \log^{O(k)} \Delta$, and that the running time of the above procedure is also $\log^{O(k)} \Delta$.
It is left to show that for every $k$-sparse $\mu'$ such that $\EMD(\mu , \mu') \leq R$,
there exists $\mu'' \in \Mc$ with $\EMD(\mu'' , \mu') \leq R / 4$.

\begin{claim}
    There exists an optimal flow between $\mu$ and $\mu'$ that is supported on at most $2k$ pairs of points.
\end{claim}
\begin{proof}
    Consider an optimal flow from $\mu$ to $\mu'$. Consider an undirected graph $G=(V,E)$ with 
$V=\supp(\mu)\cup \supp(\mu')$. We connect two vertices $(x,y) \in \supp(\mu)$ and $(x',y') \in \supp(\mu')$
iff there non-zero amount flowing from $(x,y)$ to $(x',y')$ in the flow.

If $|E|\geq 2k+1$, then there is a cycle $e_1, e_2, ..., e_{2m} \in E$ of even length $2m$ in the graph $G$. 
W.l.o.g. assume that the total length of $e_i$ with even $i$ is at most the total length of $e_i$ with odd $i$.
Let us increase all flows over $e_i$ with even $i$ and decrease all flows over $e_i$ with odd $i$
by the same amount such that at least one edge carries zero flow. Clearly, the total cost can only decrease.

Repeating the above process several times, we arrive at a flow supported on at most $2k$ edges. \end{proof}

Thus, in our enumeration algorithm at least one $c(x, y)$ corresponds to the number of outgoing flow edges
from $(x, y) \in \supp \mu$ (Line 6).
When we enumerate $l$ there is at least one choice that guesses all the lengths of the corresponding edges
within a multiplicative factor of  $1.01$ (Line 8).
Thus, there exists a measure $\widetilde{\mu}$ (not necessarily $k$-sparse) such that
\begin{itemize}
    \item $\supp \widetilde{\mu} \subseteq \supp \mu \cup \set{p(x, y, i)}_{(x, y, i) \in \Ic}$;
    \item $\EMD(\mu' , \widetilde{\mu}) \leq R / 50$;
    \item there exists a flow between $\mu$ and $\widetilde{\mu}$ of cost at most $1.02 \cdot R$
    that transports mass from a point $(x, y)$ to
    $\set{(x, y)} \cup \set{p(x, y, i)}_{i: (x, y, i) \in \Ic}$ for every $(x, y) \in \supp \mu$.
\end{itemize}



We have that $\widetilde{\mu}$ covers $\mu'$, i.e, that $\EMD(\mu',\widetilde{\mu})\leq R/100$, and that
the procedure in the pseudocode will guess $\supp(\widetilde{\mu})$ but not necessarily
$\widetilde{\mu}$. After guessing the support (Line 9), the pseudocode proceeds by trying to guess
the measure at the support (Line 10). We will show that there will be guess $\mu''$ 
made by the pseudocode with $\supp(\mu'')\subseteq\supp(\widetilde{\mu})$ 
that satisfy $\EMD(\widetilde{\mu},\mu'')\leq 2R/50$.

Fix $(x, y) \in \supp \mu$. We show to  deal with the multi-set
$\{(x, y)\} \cup \{ p(x, y, i):  (x, y, i) \in \Ic \}$.
We round down the mass in $\widetilde{\mu}$ at the coordinates $\set{p(x, y, i)}$ to the closest element of
$\set{0, m_0, 1.01 \cdot m_0, 1.01^2 \cdot m_0^2, \ldots, \min(1, R)}$ (Line 11). 
Let $\mu''$ be the resulting measure.
We also set $\mu''((x,y)):=\sum_{i:(x,y,i) \in \Ic} (\widetilde{\mu}(p(x,y,i))-\mu''(p(x,y,i)))$ (Line 19).
One can observe that   $\mu''$ is included in the set measures enumerated by our algorithm.

We now show that $\EMD(\mu'' , \widetilde{\mu})\leq 2R/50$.
The cost of $\EMD(\mu'' , \widetilde{\mu})$ comes from two sources:
\begin{enumerate}
\item Contribution from $(x,y,i) \in \Ic$ for which $\widetilde{\mu}(p(x, y, i)) < m_0$. Then $\mu''(p(x,y,i))=0$. 
There are at most $2k$ such $(x,y,i) \in \Ic$.
But we can reroute these small masses with cost at most $2 \Delta k m_0 \leq 0.02 R$.
\item
Contribution from $(x,y,i) \in \Ic$ for which $\widetilde{\mu}(p(x, y, i)) \geq m_0$.
This implies that the value of $\widetilde{\mu}(p(x, y, i))$ is within  $1\%$ of $\mu''(p(x,y,i))$.
Therefore, the total contribution of such coordinates $(x,y,i) \in \Ic$ is at most $0.01 \cdot \EMD(\mu , \widetilde{\mu}) \leq 0.02 R$.
\end{enumerate}

Thus, overall we have  $
    \|\mu' - \mu''\|_{\EMD} \leq \|\mu' - \widetilde{\mu}\|_{\EMD} +
    \|\widetilde{\mu} - \mu''\|_{\EMD} \leq (0.02 + 0.02 + 0.02) \cdot R < R / 4.
$ \end{proof}

%% file: appls.tex

%% file: acks.tex
%

%% file: appendix.tex
\section{Proofs from Section \ref{s:upper}}
\label{s:proofs}

\begin{proof}[Proof of Claim~\ref{key_estimate}]
    Let $y' \in N_i$ be a point such that $\rho(y^*, y') \leq r_i$ (recall that $N_i$ is an $r_i$-net of $Y$).
    Clearly, it is sufficient to prove that $y' \in S_i$.
    This is equivalent to the condition $\rho(y', y_{i - 1}) \leq
    \beta r_i$.
    Let us verify the latter:
    \begin{multline*}
        \rho(y', y_{i - 1}) \leq
        K \cdot \bigl(\rho(y^*, y') + \rho(y^*, y_{i - 1})\bigr) \leq
        K \cdot \bigl(r_i + \rho(y^*, y_{i - 1})\bigr) \leq
        K^2 \cdot \bigl(r_i + \rho(x, y^*) + \rho(x, y_{i - 1})\bigr) \leq \\ \leq
        K^2 \cdot \bigl(r_i + 2 \rho(x, y_{i - 1})\bigr) \leq
        K^2 \cdot \bigl(r_i + 2 q(y_{i - 1})\bigr) \leq
        K^2 \cdot (r_i + 2 \gamma r_{i - 1}) =
        K^2 \cdot (\alpha + 2 \gamma) \cdot r_{i - 1} \leq
        \alpha \beta r_{i - 1} = \beta r_i,
    \end{multline*}
    where the third inequality follows from the definition of $y'$, 
    the fourth inequality follows from the definition
    of $y^*$, the fifth inequality follows from~(\ref{distortion}),
    the sixth step follows from the statement of the Claim,
    and the penultimate step follows from~(\ref{technical_condition}).
\end{proof}

\begin{proof}[Proof of Claim~\ref{return_yi}]
    First, observe that by~(\ref{distortion}) and the fact that the algorithm returns $y_{i-1}$
    we have
    $\rho(x, y_{i - 1}) \leq
    q(y_{i - 1}) \leq \gamma r_{i - 1}$.
    Second, by~(\ref{distortion}) and Claim~\ref{key_estimate},
    $$
        \frac{\gamma r_i}{D} < \frac{q(y_i)}{D} \leq
        \rho(x, y_i) = \rho(x, S_i) \leq K \cdot \bigl(\rho(x, y^*) + \rho(y^*, S_i)\bigr)
        \leq K \cdot \bigl(\rho(x, y^*) + r_i\bigr).
    $$
    Thus,
    $$
        \rho(x, y^*) > \Bigl(\frac{\gamma}{DK} - 1\Bigr)\cdot r_i.
    $$
    Overall,
    $$
        \frac{\rho(x, y_{i - 1})}{\rho(x, y^*)} <
        \frac{\gamma r_{i - 1}}{\left(\frac{\gamma}{DK} - 1\right) \cdot r_i} = \frac{D K \gamma}{\alpha (\gamma - DK)}.
    $$
\end{proof}

\begin{proof}[Proof of Claim~\ref{return_yl}]
    If the algorithm returns $y_L$, then
    $$
        \rho(x, y_L) \leq \gamma r_L =
        2 \gamma \alpha^L \Lambda,
    $$
    where the second step follows from the definition of $r_i$.
\end{proof}

\section{Lower Bound on Measurement Complexity}
\label{s:lower}
We use $a \lesssim b$ to denote that there exists a universal constant
$C$ such that $a \leq C b$.  We use $a \gtrsim b$ to denote $b
\lesssim a$ and $a \eqsim b$ to denote $a \lesssim b \lesssim a$.

We work with the linear sparse recovery scheme as in the Introduction
(Equation \eqref{e:approx}).  We set the probability of error to be
$P={1 \over 4}$.

The following lemma generalizes the result
in~\cite{do2010lower} to general norms and nonnegative inputs.
\begin{lemma}\label{l:packinglower}
  Consider any norm $\norm{\cdot}$ over $\R^n$ for which
  $\frac{1}{n^c} \leq \frac{\norm{x}}{\norm{x}_2} \leq n^c$ for some
  constant $c$.  Further suppose that there exists a set $X \subset
  [0, \infty)^n$ of $k$-sparse vectors such that $\norm{x} \eqsim 1$
  for all $x \in X$ and $\norm{x - x'} \gtrsim 1$ for all $x \neq x'
  \in X$.  Then any linear sparse recovery scheme 
  with approximation factor $K$ over $[0, \infty)^n$
  must use $m \gtrsim {\log \abs{X} \over \log K}$ linear measurements.
\end{lemma}
\begin{proof} We first show a set of assumptions we can make without
  loss of generality, then give an algorithm to solve augmented
  indexing using sparse recovery, then analyze the algorithm.
  
  \paragraph{WLOG assumptions and setup.} First, we show that we can assume that
  $x\in X$ have coordinates that are multiples of $1/n^{c+1}$.  Let $x'$ be
  $x$ rounded to the nearest multiple of $1/n^{c+1}$ in each
  coordinate, so $\norm{x - x'}_\infty \leq 1/n^{c+1}$.  Therefore
  $\norm{x - x'}_2 \leq \sqrt{n}/n^{c+1}$ or $\norm{x - x'} \leq
  1/\sqrt{n}$.  This means that replacing $x$ with $x'$ would also
  satisfy the conditions with negligibly worse constants and have
  coordinates that are multiples of $1/n^{c+1}$.

  We would like to give a lower bound for all randomized sparse
  recovery schemes that work for each input with 3/4 probability.  By
  Yao's minimax principle, it suffices to give an explicit
  distribution on inputs for which no deterministic sparse recovery
  scheme $(A, \mathcal{A})$ can work with $3/4$
  probability. Furthermore, we may assume that $A \in \R^{m \times n}$
  has orthonormal rows (otherwise, if $A = U\Sigma V^T$ is its
  singular value decomposition, $\Sigma^{+}U^TA$ has this property and
  the transformation can be inverted before applying the algorithm).

  We use the following lemma from~\cite{do2010lower}:
  \begin{lemma}\label{lem:roundingFix}
    Consider any $m \times n$ matrix $A$ with orthonormal rows. Let $A'$
    be the result of rounding $A$ to $b$ bits per entry. Then for any $v
    \in \mathbb{R}^n$ there exists an $s \in \mathbb{R}^n$ with $A'v =
    A(v-s)$ and $\norm{s}_1 < n^2 2^{-b} \norm{v}_1$.
  \end{lemma}
  \begin{proof}
    Let $A'' = A - A'$ be the roundoff error, so each entry of $A''$
    is less than $2^{-b}$.  Then for any $v$ and $s = A^TA''v$, we
    have $As = A''v$ and
    \[
    \norm{s}_1 = \norm{A^TA''v}_1 \leq \sqrt{n}\norm{A''v}_1 \leq
    m\sqrt{n}2^{-b}\norm{v}_1 \leq n^22^{-b}\norm{v}_1.
    \]
  \end{proof}

  Now, let $A'$ be $A$ rounded to $c'\log n$ bits per entry for $c'$
  to be chosen later.  By Lemma~\ref{lem:roundingFix}, for any $v$ we
  have $A'v = A(v-s)$ for some $s$ with $\norm{s}_1 \leq n^22^{-c'\log
    n}\norm{v}_1$, so
  \[
  \norm{s} \leq n^{2c + 2-c'} \norm{v}.
  \]

  We are now ready to construct the lower bound of $m$ via a reduction
  from the one-way augmented indexing problem in communication
  complexity.  In this problem, Alice has a bit string $b$ of length
  $r \log \abs{X}$ for $r = \log n$, and Bob has an index $i^* \in [r
  \log \abs{X}]$ as well as $b_1, \dotsc, b_{i^*-1}$.  Alice must send
  a message to Bob, who must output $b_{i^*}$ with $2/3$ probability.
  It is known that the message must contain $\Omega(r \log \abs{X}) =
  \Omega(\log n \log \abs{X})$ bits.  We will show a way to use the
  sparse recovery algorithm to solve augmented indexing with $O(m \cdot \log
  n\cdot \log K)$ bits, giving the lower bound of $m \gtrsim {\log \abs{X} \over \log K}$.

  \paragraph{Algorithm to solve augmented indexing.}
  Alice turns her $r \log \abs{X}$ bits into a list $x_1, \dotsc, x_r
  \in X$.  She then defines
  \[
  z = \sum_{i=1}^r x_i / (KC)^i
  \]
  for a sufficiently large constant integer $C$ to be specified later,
  and
  \[
  y = A'z.
  \]
  Since $\norm{z} \leq \sum_{i=1}^r \norm{x_i}/(KC)^i \lesssim 1$, we
  have that $y = A(z - s)$ for some $s$ with $\norm{s} \leq
  n^{2c+2-c'}$.  Alice then sends $y$ to Bob.

  Transmitting $y$ takes $O(m \cdot \log n \cdot \log K)$ bits.  To see this, note that
  each coordinate of $z$ is a multiple of $\frac{1}{n^{c+1}(KC)^r}$ that
  is at most $n^c$, and each coordinate of $A'$ is a multiple of
  $1/n^{c'}$ that is at most $1$.  Hence each coordinate of $y=A'z$ is a
  multiple of $\frac{1}{n^{c'+c+1}(KC)^r}$ that is at most $n^{c+1}$,
  which can be represented in $\log (n^{c'+2c+2}(KC)^r) \lesssim \log n \cdot(c'+ \log K)$
  bits.  There are $m$ coordinates, so transmitting $y$ takes $O(m \cdot
  \log n \cdot (c'+\log K))$ bits.

  Now, based on his inputs $b_1, \dotsc, b_{i^*-1}$ and $i^*$, Bob can
  figure out $x_1, \dotsc, x_{i'-1}$ and wants to figure out $x_{i'}$
  for $i' = 1+\floor{i^*/\log |X|}$.  Once he learns $y = A'z = A(z -
  s)$, Bob chooses $u \in [0, \frac{1}{KCn^{c+1}}]^n$ uniformly at
  random, and computes
  \[
  y' = (KC)^{i'}(y - A\sum_{i=1}^{i'-1} x_i / (KC)^i) + Au.
  \]
  Bob then performs sparse recovery using $\mathcal{A}$ on $y'$
  getting a result $\wh{x}$.  He rounds $\wh{x}$ to the $x \in X$
  minimizing $\norm{x - \wh{x}}$.  We will show that $x = x_{i'}$ with
  at least $2/3$ probability; if this happens, Bob can recover
  $b_{i^*}$ from the associated vector $x_{i'}$.

  \paragraph{Analysis of algorithm.}
  We have that $y' = A(z' - s + u)$ for $z', s$ with $\norm{s}
  \lesssim n^{2c + 2-c'}$ and
  \[
  z' = x_{i'} + \sum_{j=1}^{r-i'} x_{i' + j} / (KC)^j =: x_{i'} + w
  \]
  for $w = \sum_{j=1}^{r-i'} x_{i' + j} / (KC)^j$ having $\norm{w}
  \lesssim 1/(KC)$.  Then $y' = A(x_{i'} + w + u - s)$.

  For now, pretend that Bob performed sparse recovery on $A(x_{i'} + w
  + u)$ instead of $A(x_{i'} + w + u - s)$.  The distribution of
  $x_{i'} + w + u$ depends on the distribution of inputs to the
  augmented indexing problem, but it is independent of the choice of
  $A$ and is over $[0, \infty)^n$.  Therefore we can choose our $A$ to
  be a matrix that lets us perform sparse recovery with $3/4$
  probability over this distribution.  Then the result $\wh{x}$ of
  sparse recovery satisfies
  \begin{equation} \label{eq1}
  \norm{\wh{x} - (x_{i'} + w + u)} \lesssim K\min_{k\text{-sparse } \overline{x}} \norm{x_{i'} + w + u - \overline{x}} \leq K\norm{w + u}
  \end{equation}
  with $3/4$ probability, or
  \begin{equation} \label{eq2}
  \norm{\wh{x} - x_{i'}} \lesssim K(\norm{w} + \norm{u}) \lesssim 1/C
  \end{equation}
  If $C$ is a sufficiently large constant, this is less than 
	\begin{equation} \label{eq3}
		\min_{x
    \neq x' \in X} \norm{x - x'}/2 \gtrsim 1.
	\end{equation}
	Therefore, when Bob
  rounds $\wh{x}$ to $X$, he gets $x_{i'}$ whenever sparse recovery
  succeeds, as happens with $3/4$ probability.

  In fact, Bob performs sparse recovery on $A(x_{i'} + w + u - s)$ not
  $A(x_{i'} + w + u)$.  However, the latter is statistically close to
  the former.  In particular, $\norm{s}_\infty \lesssim n^{3c + 2 -
    c'}$ so that the total variation distance
  \[
  \text{TV}(u, u-s) \lesssim n \cdot \frac{n^{3c + 2 - c'}}{1/(Kn^{c+1})} \leq Kn^{4c+4-c'}
  \]
  Setting $c' = 4c+5+{\log K \over \log n}$, we get that
  \[
  \text{TV}(A(x_{i'} + w + u), A(x_{i'} + w + u - s)) \leq \text{TV}(u, u-s) \lesssim 1/n.
  \]
  Therefore Bob's rounding of $\wh{x}$ to $X$ will equal $x_{i'}$ with
  probability at least $3/4 - O(1/n) > 2/3$.  This solves the
  augmented indexing problem with only $O(m \log n \cdot (c'+\log K))=O(m\log n \cdot \log K)$ bits of
  communication.  Since augmented indexing requires $\Omega(r \log
  \abs{X}) = \Omega(\log n \log \abs{X})$ bits of communication in
  this setting, we have $m \gtrsim {\log \abs{X} \over \log K}$.
\end{proof}

\begin{proof}[Proof of Theorem~\ref{t:lower}.]
  Define $S = [0, \infty)^n \cap T_k \cap \{x \in \R^n : \norm{x} \leq
  1\}$.

  Because the space and the norm are homogeneous, we have by the
  definition of doubling dimension that covering $S$ requires $2^d$
  balls of radius $1/2$.  Therefore we can find a packing $X \subset
  S$ of $2^d$ points such that $\min_{x \neq x' \in X} \norm{x - x'}
  \geq 1/2$.  This also means at most one $x \in X$ has $\norm{x} <
  1/4$.  Throwing this possible element out, we get a set of size
  $2^d-1$ satisfying the constraints of
  Lemma~\ref{l:packinglower}, giving that $m \gtrsim {\log (2^d -
  1) \over \log K} \gtrsim {d \over \log K}$.
\end{proof}

\subsection{Lower bound for streaming algorithms}

In this section we show a lower bound on the space bit complexity of any streaming algorithm
that maintains an approximately best $k$-sparse approximation of a vector with respect to any norm
$\|\cdot\|$ on $\Rbb^n$ such that
\begin{equation}
  \label{aratio_eq}
n^{-O(1)} \leq \frac{\|x\|}{\|x\|_2} \leq n^{O(1)}
\end{equation}
for every $x \in \Rbb^n$.

\begin{theorem}
  \label{streaming_lb}
  Suppose that there is an algorithm that can maintain a vector $x \in \Rbb^n$ under updates of the form
  $x_i := x_i + \delta_i$, where $\delta_i \in \Zbb$, and, moreover, suppose that we are promised that
  all entries of $x$ at any moment of time are integers between $0$ and $n^{O(1)}$. In the end,
  the algorithm is required to output a vector $y$ such that
  $$
  \|x - y\| \leq K \cdot \min_{\mbox{$k$-sparse $x^*$}} \|x - x^*\|,
  $$
  where $K > 2$ is some approximation factor. Then, the space bit complexity of the algorithm must be at least
  $$
  \Omega\left(d\cdot \frac{\log n}{\log K}\right),
  $$
  where $d$ is the doubling dimension of the non-negative $k$-sparse vectors under $\|\cdot\|$.
\end{theorem}

The rest of this section is devoted to proving this Theorem. We roughly follow the above argument for proving the lower bound for sparse recovery. However,  in this case, the argument is even simpler since we do not need to handle issues related to the sketching matrix. 

First, we take $r = 2^{\Omega(d)}$ non-negative $k$-sparse vectors $v_1, \ldots, v_r$ whose $\|\cdot\|$-norm is $\Theta(1)$ and that are $\Omega(1)$-pairwise separated wrt $\|\cdot\|$. We will show how Alice and Bob can solve Augmented Indexing on
$b = \Omega\left(d \frac{\log n}{\log K}\right)$-bit strings using the assumed algorithm.
Alice partitions her $b$-bit sequence into blocks of length $\log r = \Omega(d)$, encodes each block in one of the $v_i$'s (denote it by $u_j$ for
$1 \leq j \leq b / \log r$), and then feeds the (properly rescaled and discretized) vector
$$
U = \sum_{j = 1}^{b / \log r} \frac{u_j}{(CK)^j},
$$
where $C > 0$ is a sufficiently large constant, to the algorithm. Bob takes over, starting from this moment, subtracts the part of $U$
that corresponds to his prefix and then uses the algorithm to recover the next $u_j$.

Overall, we have that the required space is at least $\Omega(b) = \Omega\left(d \cdot \frac{\log n}{\log K}\right)$. The only remaining
fact we need to argue about is why the accuracy $[0, n^{O(1)}]$ per entry is sufficient.
First, we use~(\ref{aratio_eq}) to claim that the polynomial in $n$ accuracy is enough to represent $v_i$'s with the required conditions.
Second, since $b / \log r \approx \log n / \log K$, we get that $U$ can be represented with accuracy polynomial in $n$.

\section{Additional Bounds on Doubling Dimension for $\EMD$}
\label{s:additional}

\subsection{Upper Bound for Measures with Bounded Granularity}

\begin{lemma}
	\label{doubling_granularity}
	Consider set $S$ of all $k$-sparse measures $\mu$ such that, for all coordinates $(x,y)$,
	$\mu(x,y)$ is equal to $i/N$ for some non-negative integer $i$ and
	the total measure of $\mu$ is $1$. 
        The set $S$, under $\EMD$, has
	doubling dimension $O(k \log \log N)$.
\end{lemma}
\begin{proof}
	Let $\Ball_{\EMD}(p,r)$ be $\EMD$ ball of radius $r$ 
	containing $k$-sparse probability measures
	of granularity $1/N$ over plane. 
	$p \in S$ is the center of the ball.
	Further down we will denote $\Ball_{\EMD}(p,1)$ by $B$.

	\paragraph{Case 1.} $|\supp(p)|=1$.
	WLOG, the entire probability mass of $p$ is at point $(0,0)$. 
	We can verify that $\mu \in B$ implies 
	$\supp(\mu) \subseteq [-100N,100N]^2$.

	Let $B'$ be the set of all probability measures $\mu$ with properties that
	$\mu$ has granularity $1/n$, $\supp(\mu)\subseteq [-100N,100N]^2$,
	all coordinates of points from $\supp(\mu)$ are of the form ${i \over 1000k}$
	for an integer $i$.
	We can verify that, for every $\mu \in B$, there exists $\mu' \in B'$ with
	$\|\mu-\mu'\|\leq {|\supp(\mu)| \over 1000k}\leq 1/1000$.

	Therefore, if we construct ${1 \over 60}$-cover $X$ of $B'$ as per 
	Lemma \ref{doubling_grid} of size $|X|=(\log N)^{O(k)}$,
	$X$ is also ${1 \over 100}$-cover of $B$ and we get the required upper bound.

	There is one issue, though. It might be that the measures from cover
	$X$ does not have granularity $1/N$.
	To deal with this, we first build ${1 \over 200}$-cover $X$ of $B'$
	according to Lemma \ref{doubling_grid}. Then, for every measure $\mu$
	from the cover, if $\mu$ is not of granularity $1/N$ and
	$\EMD$ ball of radius ${1 \over 200}$ around $\mu$ does not contain
	any measure of granularity $1/N$, we discard this measure because
	it does not cover any measure of interest. If, on the other hand, the $\EMD$ ball of radius 
	$1/200$ does contain measure $\mu'$ of granularity $1/N$,
	we replace $\mu$ with $\mu'$ in the cover. Clearly, increasing
	the radius by a factor of $2$, still covers all the previous points, i.e.,
	$\Ball_{\EMD}(\mu,1/200) \subseteq \Ball_{\EMD}(\mu',1/100)$.

	\paragraph{Case 2.} $|\supp(p)|>1$.
	We denote elements of $\supp(p)$ by $(x,y)$. Because of granularity of
	measures, if $\mu \in B$, then 
	$\supp(\mu) \subseteq \bigcup_{(x,y) \in \supp(p)}S_{(x,y)}$, where
	$$
		S_{(x,y)}=[x-100N,x+100N]\times[y-100N,y+100N]
	$$
	denotes a square in plane with side length $200N$.

	We construct a graph with vertices $S_{(x,y)}$, $(x,y) \in \supp(p)$.
	We connect two vertices if the corresponding squares have non-empty
	intersection. We consider connected components of the resulting graph.
	We want to move the connected components so that, in the end, all of them
	live inside a square of side length $10^9 N^2$ and distance between
	any two connected components is $\geq 10^5 N$. We can verify that we
	can do that.

	Let $p'$ denote the resulting measure and $(x',y') \in \supp(p')$ be the
	resulting elements of the support. We round the coordinates
	of the elements of $\supp(p')$ so that all $x'$ and $y'$ are
	of the form $i \over 1000k$ for some integer $i$. Let $p''$ be the
	measure after the rounding. 

	We can check that 
	$\EMD(p',p'')\leq {\supp(p') \over 1000k} \leq 1/1000$.
	Therefore, if we construct $1 \over 100$-cover of $\Ball_{\EMD}(p'',1.1)$,
	we get $1 \over 100$-cover of $\Ball_{\EMD}(p',1)$.

	Consider all probability measures from $\Ball_{\EMD}(p'',1.2)$
	with the property that all coordinates of elements of supports of measures
	have form $i \over 1000k$ for some integer $i$.
	We denote this set by $\Ball_{\EMD}'(p'',1.2)$.
	$1 \over 200$-cover for $\Ball_{\EMD}'(p'',1.2)$ gives 
	${1 \over 200} + {1 \over 1000} < {1 \over 100}$-cover
	for $\Ball_{\EMD}(p'',1.1)$.

	To construct $1 \over 200$-cover for $\Ball_{\EMD}'(p'',1.2)$, we start by  constructing 
	$1 \over 400$-cover of $\Ball_{\EMD}'(p'',1.2)$ by measures not necessarily having
	granularity $1 \over N$. To get measures with granularity $1 \over N$, we proceed in the same way as in
	Case 1, i.e., we consider 2 cases. If a measure in the $1 \over 400$-cover
	does not have a	measure of granularity $1 \over N$ within $\EMD$ 
	distance $1 \over 400$, then discard this measure from the cover.
	Otherwise, replace the measure with the measure that has granularity 
	$1 \over N$. We can see that the set of measures that these
	operations produce, is $2 \cdot {1 \over 400} = {1 \over 200}$-cover 
	$\Ball_{\EMD}'(p'',1.2)$ and has granularity $1 \over N$.
	From Lemma \ref{doubling_grid}, the size of the cover is $(\log N)^{O(k)}$.
	As a result, we have $1 \over 100$-cover of $\Ball_{\EMD}(p',1)$.
	All measures in the cover have granularity $1 \over N$.

	Given $1 \over 100$-cover of $\Ball_{\EMD}(p',1)$, we would
	like to construct $1 \over 100$-cover of $B$. Given that $p$ and
	all measures from the cover have granularity $1 \over N$, we
	can make the following assumption. The optimal transportation 
	of probability measure from $p$ to every measure from the cover has
	probability mass on every edge of amount $i \over N$ for some
	non-negative integer $i$. As a result, if $\mu$ is a measure
	from $1 \over 100$-cover of $\Ball_{\EMD}(p',1)$, in the optimal
	transportation of $p'$ into $\mu$ (that achieves cost $\EMD(p',\mu)$),
	$\mu$ has non-zero amount on edges to elements of $\supp(p')$
	that corresponds to at most one component of the graph. (This follows
	because the connected components are highly separated in $p'$.)
	This gives that we can move components independently. We move
	the components to their original positions (according in $p$) and
	accordingly transform measures in the cover. This gives $1\over 100$-cover
	for $B$.
\end{proof}

\subsection{Lower bound on the doubling dimension}
\label{s:lower_doubling}

\begin{lemma}
	\label{lower_bound_1D_dd}
	Weighted point-sets over $[\Delta]$ of cardinality $k$ under $\EMD$ has doubling dimension 
	$\Omega(k)\cdot \log \left(\Omega(\log {\Delta \over k})\right)$ for $k>1$.
\end{lemma}
\begin{proof}
	WLOG, we assume that $\Delta$ is an integer power of $2$. 
	By $(x,w)$ we denote a point with coordinate $x$ and weight $w$.
	Let set $A$ be a weighted point-set of size $k$. 
	For $i=1,2,3,...,k/2$, we set the $i$-th point of $A$ to be $A_i=(2i\Delta/k,2)$.
	The remaining $k/2$ points has weight $0$ and arbitrary coordinates on the line.

	Let $I=i_1,i_2,...,i_{k/2}$ for $0\leq i_j \leq \log U$ (we will later set $U=\Delta/k$), and let
	 $B_I$ be a point-set defined as
	$$
		B_I=\cup_{j=1}^{k/2}\left\{(2j\Delta/k,2-2^{-i_j}),(2j\Delta/k+2^{i_j},2^{-i_j})\right\},
	$$

	We constructed $B_I$ such that $\EMD(A,B_I)=k/2$ for all $I$.

	Consider an $\EMD$ ball of radius $k/2$ around $A$, i.e., $\Ball_{\EMD}(A,k/2)$.
	We will show that the number of $\EMD$ balls of radius $k/200$ 
	needed to cover it is $\left(\log{\Delta \over k}\right)^{{k \over 2} \cdot {9 \over 10}}/2^k$,
	which yields the result.

	Consider a $k \over 200$-cover of $\Ball_{\EMD}(A,k/2)$.
	We will show that the size of the cover must be large.
	Consider a weighted pointset $B_I$ for some $I=i_1,i_2,...,i_{k/2}$.
	Given that $B_I$ is covered by an element from the cover, 
	there must be an element $C$ from the cover with the property that at least 
	$9/10$ fraction of intervals
	$$
        [2j\Delta/k+2^{i_j}  - 2^{i_j}/10 \ , \   2j\Delta/k+2^{i_j}  + 2^{i_j}/10]
	$$ 
	(for $j=1 \ldots k/2$) contains an element from $\supp(C)$.
	We call that $C$ hits $B_I$ and the set of elements of $\supp(C)$ that is contained in 
	some interval we call the hitting set of $B_I$. 
	Otherwise, for any $C$ that does not satisfy this property, we have
	$$
		\EMD(B_I,C)> \left(1-{9 \over 10}\right)\cdot |\supp(B_I)| \cdot {1 \over 10}=
			{1 \over 10} \cdot {k \over 2} \cdot {1 \over 10}={k \over 200}.
	$$

	There are
	$$
		\left|\left\{B_I|I=i_1,i_2,...,i_{k/2}\text{ and }0\leq i_j \leq \log U\text{ for }j \in [k/2]\right\}\right|
	=\left(\log{\Delta \over k}\right)^{k \over 2}
	$$
	pointsets $B_I$ that are covered.

	Consider an element $C$ from the cover. We will show that $C$ can hit at 
	most $2^k \cdot\left(\log{\Delta \over k}\right)^{{k \over 2}\cdot \left(1-{9 \over 10}\right)}$
	sets $B_I$. This will finish the proof.

	There are at most $2^k$ subsets $D$ of $\supp(C)$ that can be a hitting set for some $B_I$.
	Every $D$ can be a hitting set for at most $\left(\log{\Delta \over k}\right)^{{k \over 2}\cdot \left(1-{9 \over 10}\right)}$
	sets $B_I$ because $|D|\geq {9 \over 10} \cdot {k \over 2}$. This finishes the proof.
\end{proof}

\begin{corollary}
	\label{lower_bound_2D_dd}
	Weighted point-sets over $[\Delta]^2$ of cardinality $k$ under $\EMD$ has doubling dimension 
	$\Omega(k)\cdot \log \left(\Omega(\log {\Delta^2 \over k})\right)$ for $k>1$.
\end{corollary}
\begin{proof}
	We want to choose a point-set $A$ and a lot of 
	``highly separated'' point-sets $B_I$ similarly as in
	\ref{lower_bound_1D_dd} with $\EMD(A,B_I)=k/4$.

	For that, we place $k/4$ points with non-zero weight on a line of length 
	$\Delta\cdot {\sqrt k \over 2}$ to construct $A$ and $B_I$s analogously as in Lemma 
	\ref{lower_bound_1D_dd}. The difference is that, instead of placing $k/2$ points, we place $k/4$ points and
	that, instead of having an interval of length $\Delta$, we have interval of length $\Delta\cdot {\sqrt k \over 2}$. 
	Then we split points of $A$ with their counterparts of $B_I$
 	into $\sqrt k/2$ consecutive sequences of points each containing $\sqrt k/2$ points.
	We put $i$-th sequence in $i\cdot {2\Delta \over \sqrt k}$-row of the grid.

	We can verify that the resulting point-sets satisfy the necessary properties.
\end{proof}

\section{Better Sketches for $\EMD$}
\label{s:sketches}

\subsection{Refined analysis of the grid embedding}

In this section we recall the embedding of $\EMD_{[\Delta]^2}$ into $\ell_1$
from~\cite{it-fire-03} (building on~\cite{c-setfr-02})
and provide the refined analysis of a variant of it,
under the assumption that we are embedding a measure that can be represented as a difference
of two non-negative measures that both sum to one such that one of them is $k$-sparse.

We state the following simple lemma without a proof.
\begin{lemma}
  \label{isometry}
  For any vector $y : [\Delta] \to \R_+$, define $\CDF(y) : [\Delta] \to \R_+$ by
  \[
  \CDF(y)_i = \sum_{j \leq i} y_j.
  \]
  Then for any $y, y' : [\Delta] \to \R_+$ with $\|y\|=\|y'\|=1$ we have
  \[
  \EMD(y, y') =  \normm{1}{\CDF(y) - \CDF(y')}.
  \]
\end{lemma}

The Cauchy distribution is continuous probability distribution with the probability density function
$1 \over \pi \left(\gamma+{x^2 \over \gamma}\right)$,
where $\gamma$ is the scale parameter.
If not otherwise specified, we will refer to a Cauchy variable as one which is drawn
from distribution with $\gamma=1$.

First, we need the following folklore claim that will be useful for us later.

\begin{claim}
    \label{cauchy_tails}
    Let $X_1$, $X_2$, \ldots, $X_n$ are (not necessarily independent) non-negative
    random variables such that 
    for every $i$ and $t > 0$ we have
    $$
        \Prb{}{X_i \geq t} \leq \frac{C}{t},
    $$
    where $C > 0$ is some constant.
    Suppose that $S = \sum_i \alpha_i X_i$, where $\alpha_i \geq 0$, $\sum_i \alpha_i = 1$.
    Then, for every $\delta > 0$ we have
    $$
        \Prb{}{S \leq O_{C, \delta}(H(\alpha))} \geq 1 - \delta,
    $$
    where $H(\alpha)$ is the entropy of the distribution over $[n]$ defined by $\alpha$. In particular,
    $H(\alpha) \leq \log_2 n$.
\end{claim}
\begin{proof}
    Let $T_1$, $T_2$, \ldots, $T_n$ be non-negative parameters to be chosen later. Denote $\Ec$ the event
    ``for every $i$ one has $X_i \leq T_i$''. Then, by the union bound,
    $$
        \Prb{}{\neg \Ec} \leq \sum_{i=1}^n \frac{C}{T_i}, 
    $$
    and for every $i$ one has $\Exp{}{X_i \mid \Ec} \leq O_C(\log T_i)$.
    Thus, by Markov inequality,
    $$
        \mathrm{Pr}\Bigl[S \leq O_{C, \delta}\Bigl(\sum_i \alpha_i \log T_i\Bigr) \Bigm| \Ec\Bigr] \geq 1 - \delta / 2.
    $$
    Thus, we are looking for $T_i$'s such that $\sum_{i=1}^n \frac{C}{T_i} \leq \delta / 2$
    and $\sum_i \alpha_i \log T_i$ is minimized. Via simple calculus, we obtain the desired inequality.
\end{proof}

Let us remind, how the embedding from~\cite{it-fire-03} of $\EMD_{[\Delta]^2}$ into $\ell_1$ works.
For the sake of exposition, let us assume that $\Delta = 2^l$ for a non-negative integer $l$.


For $s = (s_1, s_2) \in \Zbb^2$ and $0 \leq t \leq l$ we define a linear map $G_{s,t} \colon \Rbb^{[\Delta]^2}
\to \ell_1$
as follows. We first impose a grid $\Gc_{s,t}$ over $\Zbb^2$ with side length $2^t$ so that one of the corners is located in $s = (s_1, s_2)$.
Then, for a measure $\mu \in \Rbb^{[\Delta]^2}$ we define $G_{s, t} \mu \in \ell_1$ as follows:
for every square of the grid we count the total mass of $\mu$ that is located there.
Then, we define the following (linear)
embedding $G_s$ of $\Rbb^{[\Delta]^2}$ into $\ell_1$ parametrized by a shift
$s = (s_1, s_2) \in \Zbb^2$: $G_s \mu := \bigoplus_{t=0}^l 2^t \cdot G_{s,t} \mu$.

In~\cite{it-fire-03} the following properties of $G_s$ have been proved.

\begin{theorem}[\cite{it-fire-03}]
    \label{grid_vanilla}
    For every $\mu \in \EMD_{[\Delta]^2}$:
    \begin{itemize}
        \item
            for every $s = (s_1, s_2) \in \Zbb^2$, one has $\|\mu\|_{\EMD} \leq O(1) \cdot \|G_s \mu\|_1$;
        \item
            $\Exp{s \in [\Delta]^2}{\|G_s \mu\|_1} \leq O(\log \Delta) \cdot \|\mu\|_{\EMD}$.
    \end{itemize}
\end{theorem}

Then, concatenating $G_s$ for all $s \in [\Delta]^2$ one obtains a \emph{deterministic}
embedding of $\EMD_{[\Delta]^2}$ into $\ell_1$ with distortion $O(\log \Delta)$.

Now we turn to the refined analysis of the above embedding.

\begin{definition}
For $x \in \Rbb^2$ and $0 < R \leq 2\Delta$ consider an $\ell_1$-ball in the plane $B_{\ell_1^2}(x, R)$.
Suppose that we sample a shift $s = (s_1, s_2) \in [\Delta]^2$ uniformly at random.
Consider the following random variable $\Ac_{x, R}(s)$:
$$
    \Ac_{x, R}(s) := \frac{\min \left(\set{2^t \colon 0 \leq t \leq l, \mbox{the grid $\Gc_{s,t}$ does not cut the
    ball $B_{\ell_1^2}(x, R)$}} \cup \set{2^{l + 1}}\right)}{R}.
$$
In words, we are looking for the side length of the finest out of $l + 1$ grids
that does not cut the ball of interest, or $2^{l+1}$, if it does not exist.
\end{definition}

Implicit in~\cite{it-fire-03} are the following two Lemmas.

\begin{lemma}[\cite{it-fire-03}]
    \label{grid_cauchy}
    There exists $C > 0$ such that
    for every $x \in \Rbb^2$, $0 < R \leq 2 \Delta$
    and $t > 0$
    one has
    $$
        \Prb{s \in [\Delta]^2}{\Ac_{x, R}(s) \geq t} \leq \frac{C}{t}.
    $$
\end{lemma}
\begin{proof}
    One has $\Ac_{x, R}(s) \geq t$ iff the coarsest grid with side length less than $R \cdot t$
    (which is $\Theta(R \cdot t)$)
    cuts the ball $B_{\ell_1}(x, R)$.
   It can be easily verified that this probability is $O(1 / t)$.
\end{proof}

\begin{lemma}[\cite{it-fire-03}]
    \label{ball_cover}
    For every two points $x, y \in [\Delta]^2$ and uniformly random $s = (s_1, s_2) \in [\Delta]^2$
    the quantity $\|G_{s} (e_{x} - e_y)\|_1$, where $e_x$ and $e_y$ are basis vectors
    that correspond to points $x$ and $y$, respectively,
    is upper bounded by $O(1) \cdot R \cdot \Ac_{u, R}(s)$
    for every $u \in \Rbb^2$ and $0 < R \leq 2 \Delta$
    such that the ball $B_{\ell_1}(u, R)$ contains both $x$ and $y$.
\end{lemma}
\begin{proof}
    All grids that are of side length at least $R \cdot \Ac_{u, R}(s)$
    do not contribute to $\|G_s (e_x - e_y)\|_1$ by the definition of $\Ac_{u, R}$.
    All finer grids contribute towards $\|G_s (e_x - e_y)\|_1$ the geometric series,
    whose total sum can be upper bounded by $O(1) \cdot R \cdot \Ac_{u, R}(s)$.
\end{proof}

Combining Lemma~\ref{grid_cauchy}, Lemma~\ref{ball_cover} and the triangle inequality,
we obtain the following Claim, which later will be very useful
for our refined analysis of the embedding from~\cite{it-fire-03}.
Basically, we show that we can upper bound $\|G_s \mu\|_1$ for $\mu \in \EMD_{[\Delta]^2}$
using Claim~\ref{cauchy_tails}.

\begin{claim}
    \label{balls_dec}
    Suppose that $\mu$ and $\nu$ are two non-negative measures over $[\Delta]^2$ that both
    sum to one.
    Assume that the optimal transportation of $\mu$ to $\nu$ consists
    of moving mass $w_i$ from the point $x_i \in [\Delta]^2$ to the point $y_i \in [\Delta]^2$
    for $1 \leq i \leq p$.
    Let $\set{B_j = B_{\ell_1}(u_j, R_j)}_{j=1}^q$ be a collection of $\ell_1$-balls in the plane
    such that for every $1 \leq i \leq p$ there exists $1 \leq j^*(i) \leq q$
    such that both $x_i$ and $y_i$ belong to $B_{j^*}$.
    For every $1 \leq j \leq q$ define
    $$
        \widetilde{w}_j = \sum_{i \colon j^*(i) = j} w_i.
    $$
    Suppose we sample a shift $s = (s_1, s_2) \in [\Delta]^2$ uniformly at random.
    Then, the random variable
    $$
        \|G_s(\mu - \nu)\|_1 \leq \sum_{i=1}^p w_i \|G_s(e_{x_i} - e_{y_i})\|_1
    $$
    is dominated by $S = \sum_{j=1}^q \widetilde{w}_j R_j \cdot X_i$
    for some non-negative (not necessarily independent) random variables $X_1$, $X_2$, \ldots, $X_q$
    such that for every $i$ and $t > 0$ one has
    $$
        \Prb{}{X_i \geq t} \leq \frac{C}{t}
    $$
    for some absolute constant $C > 0$.
\end{claim}

Now applying Claim~\ref{cauchy_tails} we conclude the following.

\begin{claim}
    \label{balls_ent}
    Assuming the notation and conditions from Claim~\ref{balls_dec},
    we have
    $$
        \Prb{s}{\|G_s(\mu - \nu)\|_1 \leq O(1) \cdot H(\alpha) \cdot T} \geq 0.99,
    $$
    where $T = \sum_{j=1}^q \widetilde{w}_j R_j = \sum_{i=1}^p w_i R_{j^*(i)}$
    and $\alpha$ is the following distribution over $[q]$:
    $$
        \alpha_j = \frac{\widetilde{w}_j R_j}{T}.
    $$
\end{claim}

Now we state two applications of this claim that are our main goal.

\begin{lemma}
    \label{grid_sparse}
    Suppose that $\mu$ and $\nu$ are two non-negative measures over $[\Delta]^2$ that both
    sum to one and, in addition, $\mu$ has support of size at most $k$ for some $1 \leq k \leq \Delta^2$.
    Then,
    $$
        \Prb{s}{\|G_s(\mu - \nu)\|_1 \leq O(\log k + \log \log \Delta) \cdot \|\mu - \nu\|_{\EMD}}\geq 0.99.
    $$
\end{lemma}
\begin{proof}
    Suppose that $\set{x_1, x_2, \ldots, x_k} \subseteq [\Delta]^2$ is the support of $\mu$. 
    Consider the following family of $O(k \log \Delta)$ balls: $\set{B(x_i, 2^j)}_{1 \leq i \leq k, 0 \leq j \leq \log \Delta + 1}$. Next, consider the optimal transportation of $\mu$ to $\nu$.
    Every edge of length $l$ participating in this transportation can be enclosed in one of the
    balls of radius $O(l)$.
    Thus, we can apply Claim~\ref{balls_ent}
    with $T \leq O(1) \cdot \|\mu - \nu\|_{\EMD}$. It is left to upper
    bound $H(\alpha)$. In this Lemma we use a crude bound: namely, that $H(\alpha) \leq
    \log O(k \log \Delta) \leq O(\log k + \log \log \Delta)$, since the support
    of $\alpha$ is of size at most $O(k \log \Delta)$.
\end{proof}

\begin{lemma}
    \label{grid_granular}
    Suppose that $\mu$ and $\nu$ are two non-negative measures over $[\Delta]^2$ that both
    sum to one, and all the weights of $\mu$ and $\nu$ are multiples of $1 / N$, where $N \geq 1$
    is some integer.
    Moreover, assume that $\mu$ is $k$-sparse for some $1 \leq k \leq N$.
    Then,
    $$
        \Prb{s}{\|G_s(\mu - \nu)\|_1 \leq O(\log k + \log \log N) \cdot \|\mu - \nu\|_{\EMD}}\geq 0.99.
    $$
\end{lemma}
\begin{proof}
    The proof is the same as in Lemma~\ref{grid_sparse},
    but we need to upper bound $H(\alpha)$ in a slightly fancier
    way.
    Let us recall the definition of $\alpha$. For each of the $O(k \log \Delta)$ balls we compute the total
    mass transported over edges that are allocated to this ball and multiply it by the radius of the ball.
    Since all the masses are multiples of $1/N$ and for every $j \leq \log \Delta$
    we have $k$ balls of radius $2^j$, we can reformulate the question of
    upper bounding $H(\alpha)$ as follows.
    Suppose that we have a bin for every $i \in [k]$ and $j \geq 0$. Then, we put $N$ balls into these bins
    (adversarially).
    Then, for each bin indexed by $(i, j)$ we multiply the number of balls there by $2^j$ and then normalize
    the resulting numbers so that they sum to $1$. What is the upper bound of the entropy of this distribution?
    We prove that it is $O(\log k + \log \log N)$ as follows.
    Denote $j^*$ the largest $j$ such that there is $i \in [k]$ such that the bin $(i, j)$ is non-empty.
    Then, the bins with $j \leq j^* - 100 \log N$ contribute to the entropy negligibly, since
    we multiply the number of balls in these bins by $2^j \leq 2^{j^*} / N^{100}$.
    But the entropy for bins with $j \geq j^* - 100 \log N$ is $\log O(k \log N) = O(\log k + \log \log N)$, since
    the total number of these ``important'' bins is $O(k \log N)$.
\end{proof}

\paragraph{Remark:}  The terms $O(\log \log \Delta)$ and $O(\log \log N)$ in Lemma~\ref{grid_sparse} and
Lemma~\ref{grid_granular},
might appear to be unfortunate artifacts of our analyses.
However, one can show that in both cases the bounds for the embedding from~\cite{it-fire-03} are \emph{in fact tight}.
Nevertheless, in the next section we show how to achieve approximation $O(\log k)$,
if we allow embeddings into more complex spaces (that still allow reasonably good sketches).

\subsection{Embedding of $\EMD$ into the $\ell_1$-sum of the small $\EMD$ instances}

In this section we provide a refined analysis of the embedding of
$\EMD_{[\Delta]^2}$
from~\cite{i-nltcf-07}.

Suppose our goal is to sketch $\EMD_{[\Delta]^2}$, where $\Delta = 2^l$ for some integer $l \geq 0$.
Let $0 \leq t \leq l$ be a parameter to be chosen later.
Let us impose a randomly shifted hierarchy of nested
grids with side lengths $\Delta / 2^t$, $\Delta / 2^{2t}$,
\ldots, $1$ ($O \big(\tfrac{\log \Delta}{t}\big)$ grids in total). By ``randomly shifted''
we mean that the coarsest
grid has a corner in a point $s = (s_1, s_2) \in [\Delta]^2$ chosen uniformly at random, and all
the finer grids are imposed by subdividing the cruder ones.
Now let us define the sketching procedure.
First, we sketch $\EMD_{[\Delta / 2^t]^2}$
instances induced by the crudest grid recursively (we have $O(2^{2t})$ of these).
Second, for each of these instances we remember the total mass.
Now, to estimate $\EMD$, we estimate $\EMD$ for the smaller instances, add these estimates,
then compute $\EMD$ for the instance induced by the total masses we remembered, multiply it by
$\Delta / 2^t$ (the side length of the crudest grid), and add it to the result.
This can be seen as a randomized embedding $f_{s} \colon \EMD_{[\Delta]^2} \to \ell_1(\EMD_{[O(2^t)]^2})$.
In~\cite{i-nltcf-07} the following properties of $f_s$ are shown:

\begin{theorem}{\cite{i-nltcf-07}}
    \label{coarse_grid}
    For every $\mu \in \EMD_{[\Delta]^2}$:
    \begin{itemize}
        \item for every $s$, one has $\|\mu\|_{\EMD_{[\Delta]^2}}
        \leq O(1) \cdot \|f_s \mu\|_{\ell_1(\EMD_{[O(2^t)]^2})}$;
        \item $\Exp{s}{\|f_s \mu\|_{\ell_1(\EMD_{[O(2^t)]^2})}}
        \leq O\Big(\frac{\log \Delta}{t}\Big) \cdot \|\mu\|_{\EMD_{[\Delta]^2}}$.
    \end{itemize}
\end{theorem}

In what follows we improve upon the second item in the above theorem under the following additional assumptions on $\mu$. 
Namely, suppose we apply $f_s$ for random $s$ to a difference $\nu - \tau$, where $\nu$ and $\tau$
are non-negative measures over $[\Delta]^2$ that sum to one and $\nu$ is $k$-sparse.

\begin{lemma}
    \label{coarse_grid_sparse}
    If $\nu$ and $\tau$ as above, then
    $$
        \Prb{s}{\|f_s (\nu - \tau)\|_{\ell_1(\EMD_{[O(2^t)]^2})}
        \leq O\Bigg(1 + \frac{\log k + \log \log \Delta}{t}\Bigg) \cdot \|\nu - \tau\|_{\EMD_{[\Delta]^2}}} \geq 0.99.
    $$
\end{lemma}
\begin{proof}
    As in the proof of Lemma~\ref{grid_sparse}, we cover the edges of optimal transportation of $\nu$ to $\tau$
    with $O(k \log \Delta)$ balls $\set{B_j}$ such that every edge of length $r$ lies within a ball of radius $O(r)$.
    Define the event $\Ec$ as follows: ``every ball $B_j$ is not cut by a grid with side length
    at least radius of $B_j$ times $C k \log \Delta$''. We can choose $C$ such that $\Prb{s}{\Ec} \geq 0.999$
    (we can take the union bound over the balls $B_j$ and for every fixed ball we proceed as in
    Claim~\ref{grid_cauchy}).

    Now let us consider a fixed edge of length $r$ from the optimal transportation.
    The goal is to argue that, conditioned on $\Ec$, the expected contribution of the edge
    to $\|f_s(\nu - \tau)\|_{\ell_1(\EMD_{[O(2^t)]^2})}$ is
    $$
        O\Bigg(r \cdot \Bigg(\frac{\log k + \log \log \Delta}{t} + 1\Bigg)\Bigg).
    $$
    Then we will be done by the triangle inequality, Markov's inequality and the fact
    that $\Prb{s}{\Ec} \geq 0.999$.

    Let us argue about the contribution of the edge for every grid separately.
    First, all grids with side length less than $r / 10$ contribute at most $O(r)$ in total,
    because the endpoints end up in different subproblems, and thus the contribution
    is proportional to the side length. The side lengths accumulate as geometric series,
    so we have that the sum is $O(r)$ in total.

    Grids with side length at least $C' \cdot r \cdot k \log \Delta$ (with $C'$ being large enough)
    do not contribute anything, conditioned
    on $\Ec$.

    Grids with side lengths between $r / 10$ and $C' \cdot r \cdot k \log \Delta$ contribute
    in expectation $O(r)$ each (see Lemma~3.3 in \cite{i-nltcf-07}). Conditioning on $\Ec$
    can change the expectation by at most a constant factor, since $\Prb{s}{\Ec} \geq 0.999$.
    Since we have $O((\log k + \log \log \Delta) / t)$ such grids,
    the required bound follows.
\end{proof}

\subsection{Implications for sketching of $\EMD$}

\begin{theorem}
	\label{sketch_implication}
    One can sketch linearly $\EMD_{[\Delta]^2}$ for measures that are differences of two non-negative measures that sum to $1$,
    one of which is $k$-sparse as follows:
    \begin{itemize}
        \item with sketch size $O(1)$ and approximation $O(\log k + \log \log \Delta)$; 
        \item with sketch size $O(\log^{\delta} \Delta)$ and approximation $O(\log k)$ for
        every \emph{constant} $0 < \delta < 1$.
        \item
    Moreover, if both measures have all the weights being multiples of $1 / n$, where $N$ is a positive integer,
    then the first of the results can be improved to having approximation $O(\log k + \log \log N)$.
    \end{itemize}
\end{theorem}
\begin{proof}
    The first result follows from composing the first item of Theorem~\ref{grid_vanilla} and
    Lemma~\ref{grid_sparse} with a sketch for $\ell_1$ from~\cite{indyksketch}.
    The third result is similar, except we use Lemma~\ref{grid_granular}.

    As for the second result, the starting point is the first item of
    Theorem~\ref{coarse_grid} together with our Lemma~\ref{coarse_grid_sparse}.
    Let us set $t = \delta \log \log \Delta$.
    This way, we get a randomized embedding of $\EMD_{[\Delta]^2}$ into $\ell_1(\EMD_{[O(\log^{\delta} \Delta)]^2})$
    with distortion $O(\log k)$.
    Then, we apply the result of Verbin and Zhang~\cite{vz-rsdre-12}
    to perform dimension reduction. Namely, we need to apply
    their randomized map twice to reduce the dimension to $O(\log \log \Delta)$. As a result, we get 
    a sketch of size $O(\log^{O(\delta)} \Delta)$ and distortion $O(\log k)$, if $\delta$
    is a (small) positive constant.
\end{proof}

\begin{theorem}
	\label{sketch_implication2}
    One can sketch linearly $\EMD_{[\Delta]}$ over interval $[\Delta]$ of measures that are differences of two non-negative measures that sum to $1$,
    one of which is $k$-sparse. We can achieve sketch size $O(1/\epsilon^2)$ and approximation $1+\epsilon$.
\end{theorem}
\begin{proof}
Using Lemma \ref{isometry}, we can isometrically embed $\EMD$ over the interval $[\Delta]$ into $\ell_1$.
Now we can sketch $\ell_1$ using the sketch from \cite{indyksketch}. This give sketch size $O(1/\epsilon^2)$ and
approximation $1+\epsilon$.
\end{proof}

\section{Sparse recovery for $\EMD$}
\label{s:sremd}
	The following three theorems follow from Lemma~\ref{recovery_upper} and Theorem~\ref{sketch_implication}.

	\begin{theorem}
		\label{upperbound_square}
		There is a linear sketching scheme of probability distributions over $[\Delta]^2$ with the following guarantees.
		The size of the sketch is
		$$
			O(k(\log \log \Delta)\log(\log k+\log \log \Delta)+\log \log(\Delta/\lambda))
		$$
		and, given a sketch of $x$, we can recover $x^*$ such that
		$$
			\EMD(x,x^*)\leq \max(O(\log k+\log \log \Delta)\min_{k\text{ - sparse }x'}\EMD(x,x'),\lambda).
		$$
		in time polynomial in $\Delta$ and $\log^{O(k)} \Delta$.
	\end{theorem}
	\begin{proof}
		Lemma~\ref{doubling_grid} gives that the doubling dimension of $k$-sparse probability measures
		over $[\Delta]^2$ is $O(k\log \log \Delta)$. Combining this with
		Lemma~\ref{recovery_upper} and the first result from Theorem~\ref{sketch_implication}, we
		get the stated guarantees.
	\end{proof}

	\begin{theorem}
		There is a linear sketching scheme of probability distributions over $[\Delta]^2$ with the following guarantees.
		The size of the sketch is
		$$
			O(1)(\log^{\delta}\Delta)
			(k(\log \log \Delta)\log\log k+\log \log(\Delta/\lambda))
		$$
		for some constant $\delta>0$. Given a sketch of $x$, we can recover $x^*$ such that
		$$
			\EMD(x,x^*)\leq \max(O(\log k)\min_{k\text{ - sparse }x'}\EMD(x,x'),\lambda).
		$$
		in time polynomial in $\Delta$ and $\log^{O(k)} \Delta$.
	\end{theorem}
	\begin{proof}
		Lemma~\ref{doubling_grid} gives that the doubling dimension of $k$-sparse probability measures
		over $[\Delta]^2$ is $O(k\log \log \Delta)$. Combining this with
		Lemma~\ref{recovery_upper} and the second result from Theorem~\ref{sketch_implication}, we
		get the stated guarantees.
	\end{proof}

	\begin{theorem}
		Let $N$ be a positive integer. There is a linear sketching scheme of probability measures that
		have granularity $1/N$.
		The size of the sketch is
		$$
			O(k(\log \log N)\log(\log k+\log \log N)+\log \log(\Lambda/\lambda))
		$$
		and, given a sketch of $x$, we can recover $x^*$ such that
		$$
			\EMD(x,x^*)\leq \max(O(\log k+\log \log N)\min_{k\text{ - sparse }x'}\EMD(x,x'),\lambda).
		$$
		in time polynomial in $\Delta$ and $\log^{O(k)} \Delta$.
		$\Lambda$ is the upper bound on $\EMD(x,y)$ for the starting $k$-sparse approximation $y$ of $x$.
	\end{theorem}
	\begin{proof}
		Lemma~\ref{doubling_granularity} gives that the doubling dimension of $k$-sparse probability measures
		with granularity $1/n$ is $O(k\log \log N)$. Combining this with
		Lemma~\ref{recovery_upper} and the third result from Theorem~\ref{sketch_implication}, we
		get the stated guarantees.
	\end{proof}

	\begin{theorem}
		\label{upperbound_interval}
		There is a linear sketching scheme of probability distributions over interval $[\Delta]$ with the following guarantees.
		The size of the sketch is
		$$
			O(1/\epsilon^2)
			(k(\log \log \Delta)\log{1 \over \epsilon}+\log \log(\Delta/\lambda))
		$$
		and, given a sketch of $x$, we can recover $x^*$ such that
		$$
			\EMD(x,x^*)\leq \max((1+\epsilon)\min_{k\text{ - sparse }x'}\EMD(x,x'),\lambda).
		$$
		in time polynomial in $\Delta$ and $\log^{O(k)} \Delta$.
	\end{theorem}	
	\begin{proof}
		Lemma~\ref{doubling_grid} also gives that the doubling dimension of $k$-sparse probability measures
		over interval $[\Delta]$ is $O(k\log \log \Delta)$. Combining this with
		Lemma~\ref{recovery_upper} and Theorem~\ref{sketch_implication2}, we
		get the stated guarantees.
	\end{proof}

\subsection{Lower Bounds for Sparse Recovery for Earth Mover's Distance}

Lemma~\ref{lower_bound_1D_dd} and Lemma~\ref{l:packinglower} gives the following two theorems \ref{lowerbound_interval} and \ref{lowerbound_square}.

\begin{theorem}
\label{lowerbound_interval}
Any linear sparse recovery scheme with approximation factor $K$ with respect to $\EMD$ over interval $[\Delta]$ requires
$$
	m\geq { \Omega(k)\log\left(\Omega(\log{\Delta \over k})\right)\over \log K} 
$$
measurements for sparsity $k>1$.
\end{theorem}

We want to compare guarantees of Theorem~\ref{upperbound_interval} with the lower bound that
we achieve in Theorem~\ref{lowerbound_interval}.

Theorem~\ref{upperbound_interval} and assumptions that
$\epsilon$ is a constant and $\lambda \geq 2^{-(\log \Delta)^{O(k)}}$ gives approximation guarantee
\begin{equation}
	\label{guarantee}
	\EMD(x,x^*)\leq \max(O(1)\min_{k\text{ - sparse }x'}\EMD(x,x'),\lambda)
\end{equation}
with $O(k \log \log \Delta)$ number of measurements.

Theorem~\ref{lowerbound_interval} and assumption that $k<\Delta^{1-c}$ for some constant $c>0$ give lower bound 
$\Omega(k \log \log \Delta)$ on the number of measurements for constant approximation factor.
However, this lower bound holds for the case when $\lambda$ is equal to $0$ in guarantee \ref{guarantee}.

From the proof of Lemma~\ref{l:packinglower} (equations \eqref{eq1}, \eqref{eq2} and 
\eqref{eq3}) and Lemma~\ref{lower_bound_1D_dd} 
(we construct ${k \over 200}$-cover for $\EMD$ ball of radius $k/2$)
we see that we are actually good as long as $\lambda$ is sufficiently small. As long as $\lambda\leq {k \over C}$
for some large constant $C$. Therefore, our lower bound holds if ${k \over C} \geq \lambda \geq  2^{-(\log \Delta)^{O(k)}}$.

We see that the upper bound and the lower bound match for the described range of parameters.

\begin{theorem}
\label{lowerbound_square}
Any linear sparse recovery scheme with approximation factor $K$ with respect to $\EMD$ over square $[\Delta]^2$ requires
$$
	m\geq { \Omega(k)\log\left(\Omega(\log{\Delta^2 \over k})\right)\over \log K} 
$$
measurements for sparsity $k>1$.
\end{theorem}

We want to compare guarantees of Theorem~\ref{upperbound_square} with the lower bound that
we achieve in Theorem~\ref{lowerbound_square}.

Theorem~\ref{upperbound_square} and assumptions that
$\epsilon$ is a constant and $\lambda \geq 2^{-(\log \Delta)^{O(k)}}$ gives approximation guarantee
\begin{equation}
	\label{guarantee2}
	\EMD(x,x^*)\leq \max(O(\log k+\log \log \Delta)\min_{k\text{ - sparse }x'}\EMD(x,x'),\lambda)
\end{equation}
with $O(k (\log \log \Delta)\log(\log k+\log \log \Delta))$ number of measurements.

Theorem~\ref{lowerbound_square} and assumption that $k<\Delta^{2-c}$ for some constant $c>0$ give lower bound 
$\Omega(1){k \log \log \Delta \over \log(\log k+\log \log \Delta)}$ on the number of measurements for approximation factor
$O(\log k+\log \log \Delta)$.
However, this lower bound holds for the case when $\lambda$ is equal to $0$ in guarantee \ref{guarantee2}.

From the proof of Lemma~\ref{l:packinglower} (equations \eqref{eq1}, \eqref{eq2} and 
\eqref{eq3}) and Corollary~\ref{lower_bound_2D_dd} 
(we construct ${k \over 200}$-cover for $\EMD$ ball of radius $k/2$)
we see that we are actually good as long as $\lambda$ is sufficiently small. As long as $\lambda\leq {k \over C}$
for some large constant $C$. Therefore, our lower bound holds if ${k \over C} \geq \lambda \geq  2^{-(\log \Delta)^{O(k)}}$.

We see that the upper bound and the lower bound match up to a factor of $\log^2(\log k+\log \log \Delta)$ 
for the described range of parameters.
\section{Sketching of $1$-Median}
\label{s:1median}

For a vector $x \in \R^n$, we use $\normm{med}{x}$ to denote the
median over $i\in [n]$ of $\abs{x_i}$.

\subsection{Subspace embeddings}

\begin{lemma}
  Let $L$ be a $d$-dimensional subspace of $\R^n$.  Let $A \in
  \R^{m\times n}$ be a matrix with $m = O(\frac{1}{\eps^2}d \log
  \frac{d}{\eps \delta})$ and i.i.d. Cauchy entries with scale
  parameter $\gamma = 1$.  With $1 - \delta$ probability, for all $x
  \in L$ we have
  \[
  (1-\eps)\normm{1}{x} \leq \normm{med}{Ax} \leq (1+\eps) \normm{1}{x}.
  \]
\end{lemma}
\begin{proof}
  In an abuse of notation, let $L$ be an orthonormal basis for the
  subspace $L$.  For any threshold $\tau = \text{poly}(\frac{d}{\eps
    \delta})$, the probability that any entry of $AL$ has absolute
  value larger than $\tau$ is $O(\sqrt{d}/\tau)$, using that the
  $\ell_1$ norms of the columns of $L$ is $\sqrt{d}$.  Setting $\tau =
  O(d^{2.5}/\delta)$, we have that every entry of $AL$ is at most
  $\tau$ with probability $1-\delta/2$.  Suppose this happens.

  Then for all $x \in \R^d$, we have that $\normm{1}{Lx} \geq
  \normm{2}{Lx} = \normm{2}{x} \geq \normm{1}{x}/\sqrt{d}$ and
  $\normm{\infty}{ALx} \leq \tau \normm{1}{x} \leq \tau
  \sqrt{d}\normm{1}{Lx}$.  Thus for all $y \in L$ we have
  \[
  \normm{\infty}{Ay} \leq (d^3/\delta) \normm{1}{y}.
  \]
  Let $\tau' = d^3/\delta$.

  We construct an $\frac{\eps}{\tau'}$-net $T$ in the $\ell_1$ norm for
  the unit $\ell_1$ ball intersect $L$, which has size at most $(1 +
  \tau'/\eps)^d = e^{O(d \log \frac{d}{\eps \delta})}$ by the standard volume argument.

  For any $x \in \R^n$, we say $Ax$ is ``good'' if only a
  $\frac{1}{2}-C_2\eps$ fraction of coordinates are too large or too
  small, i.e.
  \begin{align*}
    \abs{\{i \st \abs{(Ax)_i} < (1-\eps) \normm{1}{x} \}} &\leq (\frac{1}{2} - C_2\eps )m\\
    \abs{\{i \st \abs{(Ax)_i} > (1+\eps) \normm{1}{x} \}} &\leq
    (\frac{1}{2} - C_2\eps )m
  \end{align*}
  for some small constant $C_2$.  If $Ax$ is ``good'', then for any
  $y$ with at most $C_2 \eps m$ coordinates larger than $\eps
  \normm{1}{x}$, we have
  \begin{align}
    (1-2\eps) \normm{1}{x} \leq \normm{med}{Ax + y} \leq
    (1+2\eps)\normm{1}{x}.\label{eq:medadjust}
  \end{align}

  Because $(Ax)_i$ is a Cauchy variable with scale $\normm{1}{x}$, we
  have that
  \begin{align*}
    \Pr[\abs{(Ax)_i} < (1-\eps)\normm{1}{x}] &< 1/2 - \Omega(\eps)\\
    \Pr[\abs{(Ax)_i} > (1+\eps)\normm{1}{x}] &< 1/2 - \Omega(\eps).
  \end{align*}
  By a Chernoff bound, for sufficiently small $C_2$ we have that $Ax$
  is ``good'' with all but $e^{-\Omega(\eps^2 m)}$ probability.  For our choice of $m$, we can
  union bound to have that $Ax$ is ``good'' for all $x \in T$ with all
  but $e^{-\Omega(\eps^2 m)} \leq \delta^{\Omega(d)}$ probability.

  Every $y \in L$ with $\normm{1}{y} = 1$ can be expressed as $x + z$
  for $x \in T$ and $\normm{1}{z} \leq \eps/\tau'$.  We have that $Ax$
  is ``good'' and that $\norm{\infty}{Az} \leq \tau' \normm{1}{z} \leq
  \eps$.  Hence by~\eqref{eq:medadjust},
  \[
  (1-2\eps) \normm{1}{x} \leq \normm{med}{Ay} \leq (1+2\eps)\normm{1}{x}.
  \]
  which implies
  \[
  (1-3\eps) \normm{1}{y} \leq \normm{med}{Ay} \leq (1+3\eps)\normm{1}{y}.
  \]
  Since $A$ is linear, the restriction to $\normm{1}{y} = 1$ is
  unnecessary; rescaling $\eps$ then gives the result.
\end{proof}

\begin{corollary}\label{cor:spacemin}
  Let $A$ have $O(d\log (d/(\eps\delta)) / \eps^2)$ rows and Cauchy
  entries with scale $\gamma = 1$.  For any subspace $L$ of dimension
  $d$ and subset $S \subset L$, with $1-\delta$ probability we
  have that
  \[
  \wh{x} := \argmin_{x \in S} \normm{med}{Ax}
  \]
  satisfies
  \[
  \normm{1}{\wh{x}} \leq (1 + \eps) \min_{x \in S} \normm{1}{x}.
  \]
\end{corollary}

\subsection{$1$-median in $d$ dimensions}

\subsubsection{$1$-median in $1$ dimension}

\begin{theorem}
  We can find a $1+\eps$-approximation to the $1$-median in $1$
  dimensions using $O(\log(1/\eps)/\eps^2)$ linear measurements and
  $\exp(\poly(1/\eps))$ time.
\end{theorem}
\begin{proof}
  Define $B \in \R^{n \times 2}$ by $B_{i, 1} = i$ and $B_{i, 2} = 1$
  for all $i \in [n]$.  For any $x \in \R^n$, define $D_x \in
  \R^{n\times n}$ to be the diagonal matrix with $D_{i, i} = x_i$.
  Then for any $j \in [n]$ and $z = (j, -1)$, we have that
  \[
  \normm{1}{D_x B z} = \sum_i \abs{x_i}\abs{i - j}
  \]
  is the cost of using $j$ as the median for $x$.

  Let $A \in \R^{m \times n}$ for $m = O(\log(1/\eps)/\eps^2)$ have
  i.i.d. Cauchy entries.  Then $AD_xB \in \R^{m \times 2}$ consists of
  $2m$ linear measurements of $x$.

  Furthermore, the set of $S = \{D_x B z \mid z_2 = -1\}$ is a subset
  of a 2-dimensional subspace.  Hence, by
  Corollary~\ref{cor:spacemin},
  \[
  \wh{z} = \argmin_{\substack{z \in \R^2\\z_2 = -1}} \normm{med}{AD_xBz}
  \]
  satisfies
  \[
  \normm{1}{D_xB\wh{z}} \leq (1 + \eps) \min_{\substack{z \in \R^2\\z_2 = -1}} \normm{1}{D_x B z} = (1+\eps)\text{cost}(x).
  \]
  Given $AD_xB$ we can compute $\wh{z}$, from which we recover $z_1$
  as a $(1+\eps)$ approximation to the $1$-median.
\end{proof}

\subsubsection{$1$-median in $d$ dimensions}

\begin{claim}[Dvoretsky's Theorem~\cite{dvor}]\label{c:dvoretsky}
  Let $G \in \R^{m \times d}$ have suitably scaled i.i.d. Gaussian
  entries, for $m = O(d/\eps^2)$.  Then with all but $e^{-\Omega(d)}$
  probability, for all $x \in \R^d$ we have
  \[
  \normm{1}{Gx} \leq \normm{2}{x} \leq (1+\eps)\normm{1}{Gx}.
  \]
\end{claim}

\begin{theorem}
  We can find a $1+\eps$-approximation to the Euclidean $1$-median in
  $d$ dimensions using $O(d^2\log(d/\eps)/\eps^2)$ linear measurements
  and $\exp(\poly(d/\eps))$ time.
\end{theorem}
\begin{proof}
  Let $G \in \R^{t \times d}$ for $t = O(d /\eps^2)$ satisfy
  Claim~\ref{c:dvoretsky}, so
  \[
  \normm{1}{Gp} \leq \normm{2}{p} \leq (1+\eps)\normm{1}{Gp}
  \]
  for all $p \in [n]^d$.  For each point $p \in [n]^d$, define the
  matrix $B^{(p)} \in \R^{t \times (t+1)}$ by the first $t$ columns
  being the identity matrix and column $t+1$ being $Gp$.
  
  Define $G' \in \R^{(t+1) \times (d+1)}$ to equal $G$ over the first
  $t \times d$ submatrix, $G_{t+1, d+1} = 1$, and zero elsewhere.  For
  any point $p \in [n]^d$ define $z^{(p)} \in \R^{d+1}$ by $z_i = p_i$
  for $i \leq d$ and $z_{d+1} = -1$.  For any $p, q \in [n]^d$, we
  have
  \[
  B^{(q)}G'z_p = 
\left(
    \begin{array}{cc}
     I & Gq
    \end{array}
\right)
\left(
    \begin{array}{cc}
      G & 0\\0 & 1
    \end{array}
\right)
\left(
    \begin{array}{c}
      p \\-1
    \end{array}
\right) = Gq - Gp
  \]
  Hence
  \[
  \normm{1}{B^{(q)}G'z_p} = \normm{1}{Gq - Gp} \leq \normm{2}{p-q} \leq
  (1+\eps)\normm{1}{B^{(q)}G'z_p}.
  \]
  For $x \in \R^{n^d}$, define $C_x \in \R^{tn^d \times (t+1)}$ to be
  the concatenation of the matrices $x_p B^{(p)}$ for all $p \in
  [n]^d$.  Then for all $x \in \R^{n^d}$ and $p \in [n]^d$, therefore,
  \begin{align}
    \normm{1}{C_x G' z^{(p)}} \leq \text{cost}(x, p) \leq
    (1+\eps)\normm{1}{C_x G' z^{(p)}}.
    \label{e:zpcost}
  \end{align}
  Let $A \in \R^{m \times tn^d}$ for $m = O(d\log(d/\eps)/\eps^2)$
  have i.i.d. Cauchy entries.  Our method observes
  \[
  AC_xG' \in \R^{m \times (d+1)}
  \]
  which is a set of $m(d+1) = O(d^2\log(d/\eps)/\eps^2)$ linear
  measurements of $x$.

  By Corollary~\ref{cor:spacemin}, with good probability we have that
  \[
  \wh{z} = \argmin_{\substack{z \in \R^{d+1}\\z_{d+1} = -1}} \normm{med}{AC_xG'z}
  \]
  satisfies
  \[
  \normm{1}{C_x G' \wh{z}} \leq (1 + \eps) \min_{\substack{z \in
      \R^{d+1}\\z_{d+1} = -1}} \normm{1}{C_x G'z} =
  (1+\eps)\text{cost}(x).
  \]
  Hence by~\eqref{e:zpcost}, for $\wh{p} = (\wh{z}_1, \dotsc, \wh{z}_d)$,
  \[
  \text{cost}(x, \wh{p}) \leq (1+\eps)^2 \text{cost}(x).
  \]

  Given $AC_xG'$ we can compute $\wh{z}$, from which we get $\wh{p}$
  as a $(1+\eps)$ approximation to the $1$-median.
\end{proof}

\section{Lower bounds for $k$-means}
\label{s:kmeans}
In this section we prove lower bounds for sketching and streaming \emph{$k$-means}.

First, one can extend the definition of $\EMD$ to the sum of squares of distances.
Let us denote the corresponding ``distance'' $\EMD^2$. It is immediate to see that
$\Rbb^{[\Delta]^2}$ equipped with $\EMD^2$ is a $2$-quasi-metric space.
Sparse recovery with respect to $\EMD^2$ is equivalent to the $k$-means clustering.

Second, observe that the construction from Section~\ref{s:lower_doubling} can be translated verbatim
to $\EMD^2$ to show that the doubling dimension of the latter is $\Omega(k \cdot \log \log \tfrac{\Delta^2}{k})$
as well.

Finally, observe that the results of Section~\ref{s:lower} can be applied to $\EMD^2$ as well.
Indeed, $\EMD^2$ enjoys the polynomial aspect ratio and relaxed triangle inequality,
and these two happen to be enough for the argument to go through.

As a result, we get the lower bound $\Omega(k \cdot \log \log \tfrac{\Delta^2}{k} / \log K)$
on the number of measurements necessary for the linear sketching of $k$-means with approximation $K$.

Alternatively, we can consider the streaming model and reuse the proof from Section~\ref{s:lower}
to show that \emph{streaming} $k$-means with approximation $K$ requires
$$
\Omega\left(\frac{k}{\log K} \cdot \log \log \frac{\Delta^2}{k} \cdot \log \Delta\right)
$$
\emph{bits}.

%% file: main.bbl
\newcommand{\etalchar}[1]{$^{#1}$}
\begin{thebibliography}{CDM{\etalchar{+}}13}

\bibitem[ABS10]{abs10}
Marcel~R. Ackermann, Johannes Bl{\"{o}}mer, and Christian Sohler.
\newblock Clustering for metric and nonmetric distance measures.
\newblock {\em {ACM} Transactions on Algorithms}, 6(4), 2010.

\bibitem[AZGR15]{allen2014restricted}
Zeyuan Allen-Zhu, Rati Gelashvili, and Ilya Razenshteyn.
\newblock Restricted isometry property for general p-norms.
\newblock In {\em Proceedings of 31st International Symposium on Computational
  Geometry (SoCG 2015)}, 2015.

\bibitem[BC05]{bc05}
Bo~Brinkman and Moses Charikar.
\newblock On the impossibility of dimension reduction in l\({}_{\mbox{1}}\).
\newblock {\em J. {ACM}}, 52(5):766--788, 2005.

\bibitem[BGI{\etalchar{+}}08]{BGIKS}
Radu Berinde, Anna~C. Gilbert, Piotr Indyk, Howard~J. Karloff, and Martin~J.
  Strauss.
\newblock {Combining geometry and combinatorics: A unified approach to sparse
  signal recovery}.
\newblock {\em Allerton}, 2008.

\bibitem[BI14]{bavckurs2014better}
Art{\=u}rs Ba{\v{c}}kurs and Piotr Indyk.
\newblock Better embeddings for planar earth-mover distance over sparse sets.
\newblock In {\em Annual Symposium on Computational Geometry}, page 280. ACM,
  2014.

\bibitem[CDM{\etalchar{+}}13]{cdmmmw13}
Kenneth~L. Clarkson, Petros Drineas, Malik Magdon{-}Ismail, Michael~W. Mahoney,
  Xiangrui Meng, and David~P. Woodruff.
\newblock The fast cauchy transform and faster robust linear regression.
\newblock In {\em Proceedings of the Twenty-Fourth Annual {ACM-SIAM} Symposium
  on Discrete Algorithms, {SODA} 2013, New Orleans, Louisiana, USA, January
  6-8, 2013}, pages 466--477, 2013.

\bibitem[Cha02]{c-setfr-02}
Moses Charikar.
\newblock Similarity estimation techniques from rounding algorithms.
\newblock In {\em Proceedings on 34th Annual {ACM} Symposium on Theory of
  Computing, May 19-21, 2002, Montr{\'{e}}al, Qu{\'{e}}bec, Canada}, pages
  380--388, 2002.

\bibitem[CRT06]{CRT06:Stable-Signal}
E.~J. Cand{\`e}s, J.~Romberg, and T.~Tao.
\newblock Stable signal recovery from incomplete and inaccurate measurements.
\newblock {\em Comm. Pure Appl. Math.}, 59(8):1208--1223, 2006.

\bibitem[CS02]{cs02}
Moses Charikar and Amit Sahai.
\newblock Dimension reduction in the {\textbackslash}ell {\_}1 norm.
\newblock In {\em 43rd Symposium on Foundations of Computer Science {(FOCS}
  2002), 16-19 November 2002, Vancouver, BC, Canada, Proceedings}, pages
  551--560, 2002.

\bibitem[DIPW10]{do2010lower}
Khanh {Do Ba}, Piotr Indyk, Eric Price, and David~P Woodruff.
\newblock Lower bounds for sparse recovery.
\newblock In {\em SODA}, volume~10, pages 1190--1197. SIAM, 2010.

\bibitem[Don06]{Don06:Compressed-Sensing}
D.~L. Donoho.
\newblock {C}ompressed {S}ensing.
\newblock {\em IEEE Trans. Info. Theory}, 52(4):1289--1306, Apr. 2006.

\bibitem[Dvo60]{dvor}
A.~Dvoretzky.
\newblock {Some results on convex bodies and Banach spaces}.
\newblock 1960.

\bibitem[FS05]{FSoh}
G.~Frahling and C.~Sohler.
\newblock Coresets in dynamic geometric data streams.
\newblock {\em STOC}, 2005.

\bibitem[GD05]{GD}
K.~Grauman and T.~Darrell.
\newblock The pyramid match kernel: Discriminative classification with sets of
  image features.
\newblock {\em ICCV}, 2005.

\bibitem[GIIS14]{gilbert2014recent}
A~Gilbert, P~Indyk, M~Iwen, and L~Schmidt.
\newblock Recent developments in the sparse fourier transform: A compressed
  fourier transform for big data.
\newblock {\em Signal Processing Magazine, IEEE}, 31(5):91--100, 2014.

\bibitem[GIP10]{gupta2010sparse}
Rishi Gupta, Piotr Indyk, and Eric Price.
\newblock Sparse recovery for earth mover distance.
\newblock In {\em Communication, Control, and Computing (Allerton), 2010 48th
  Annual Allerton Conference on}, pages 1742--1744. IEEE, 2010.

\bibitem[GKK10]{gkk-ecmd-10}
Lee-Ad Gottlieb, Leonid Kontorovich, and Robert Krauthgamer.
\newblock Efficient classification for metric data.
\newblock In {\em Proceedings of the 23rd Conference on Learning Theory (COLT
  2010)}, pages 433--440, 2010.

\bibitem[Ind04]{I04}
P.~Indyk.
\newblock Algorithms for dynamic geometric problems over data streams.
\newblock {\em STOC}, 2004.

\bibitem[Ind06]{indyksketch}
P.~Indyk.
\newblock Stable distributions, pseudorandom generators, embeddings and data
  stream computation.
\newblock {\em J. ACM}, 53(3):307--323, 2006.

\bibitem[Ind07]{i-nltcf-07}
Piotr Indyk.
\newblock A near linear time constant factor approximation for {E}uclidean
  bichromatic matching (cost).
\newblock In {\em Proceedings of the 18th ACM-SIAM Symposium on Discrete
  Algorithms (SODA 2007)}, pages 39--42, 2007.

\bibitem[IP11]{indyk2011k}
Piotr Indyk and Eric Price.
\newblock K-median clustering, model-based compressive sensing, and sparse
  recovery for earth mover distance.
\newblock In {\em Proceedings of the forty-third annual ACM symposium on Theory
  of computing}, pages 627--636. ACM, 2011.

\bibitem[IT03]{it-fire-03}
Piotr Indyk and Nitin Thaper.
\newblock Fast image retrieval via embeddings.
\newblock In {\em the 3rd International Workshop on Statistical and
  Computational Theories of Vision (at ICCV 2003)}, 2003.

\bibitem[JST11]{jst11}
Hossein Jowhari, Mert Saglam, and G{\'{a}}bor Tardos.
\newblock Tight bounds for lp samplers, finding duplicates in streams, and
  related problems.
\newblock In {\em Proceedings of the 30th {ACM} {SIGMOD-SIGACT-SIGART}
  Symposium on Principles of Database Systems, {PODS} 2011, June 12-16, 2011,
  Athens, Greece}, pages 49--58, 2011.

\bibitem[KL04]{kl-nnsap-04}
Robert Krauthgamer and James~R. Lee.
\newblock Navigating nets: simple algorithms for proximity search.
\newblock In {\em Proceedings of the 15th ACM-SIAM Symposium on Discrete
  Algorithms (SODA 2004)}, pages 798--807, 2004.

\bibitem[MD13]{mo2013compressive}
Dian Mo and Marco~F Duarte.
\newblock Compressive parameter estimation with earth mover's distance via
  k-median clustering.
\newblock In {\em SPIE Optical Engineering+ Applications}, pages
  88581P--88581P. International Society for Optics and Photonics, 2013.

\bibitem[MM13]{mm13}
Xiangrui Meng and Michael~W. Mahoney.
\newblock Low-distortion subspace embeddings in input-sparsity time and
  applications to robust linear regression.
\newblock In {\em Symposium on Theory of Computing Conference, STOC'13, Palo
  Alto, CA, USA, June 1-4, 2013}, pages 91--100, 2013.

\bibitem[Mut05]{muthukrishnan2005data}
S~Muthukrishnan.
\newblock {\em Data streams: Algorithms and applications}.
\newblock Now Publishers Inc, 2005.

\bibitem[RTG00]{RTG}
Y.~Rubner, C.~Tomassi, and L.~J. Guibas.
\newblock The earth mover's distance as a metric for image retrieval.
\newblock {\em International Journal of Computer Vision}, 40(2):99--121, 2000.

\bibitem[SW11]{sw11}
Christian Sohler and David~P. Woodruff.
\newblock Subspace embeddings for the l\({}_{\mbox{1}}\)-norm with
  applications.
\newblock In {\em Proceedings of the 43rd {ACM} Symposium on Theory of
  Computing, {STOC} 2011, San Jose, CA, USA, 6-8 June 2011}, pages 755--764,
  2011.

\bibitem[VZ12]{vz-rsdre-12}
Elad Verbin and Qin Zhang.
\newblock Rademacher-sketch: A dimensionality-reducing embedding for
  sum-product norms, with an application to {E}arth-{M}over {D}istance.
\newblock In {\em Proceedings of the 39th International Colloquium on Automata,
  Languages and Programming (ICALP 2012)}, pages 834--845, 2012.

\bibitem[War14]{ward2014unified}
Rachel Ward.
\newblock A unified framework for linear dimensionality reduction in $\ell_1$.
\newblock {\em arXiv preprint arXiv:1405.1332}, 2014.

\bibitem[WZ13]{wz13}
David~P. Woodruff and Qin Zhang.
\newblock Subspace embeddings and
  {\textbackslash}({\textbackslash}ell{\_}p{\textbackslash})-regression using
  exponential random variables.
\newblock In {\em {COLT} 2013 - The 26th Annual Conference on Learning Theory,
  June 12-14, 2013, Princeton University, NJ, {USA}}, pages 546--567, 2013.

\bibitem[YO14]{yousefi2014improved}
Arman Yousefi and Rafail Ostrovsky.
\newblock Improved approximation algorithms for earth-mover distance in data
  streams.
\newblock {\em arXiv preprint arXiv:1404.6287}, 2014.

\end{thebibliography}
